\documentclass[12pt]{article}

\usepackage{graphicx,mathrsfs,bbm,amssymb,here,amsmath,cite}

\newcommand{\be}{\begin{equation}}
\newcommand{\ee}{\end{equation}}
\newcommand{\bea}{\begin{eqnarray}}
\newcommand{\eea}{\end{eqnarray}}
\newcommand{\ba}{\begin{array}}
\newcommand{\ea}{\end{array}}
\newcommand{\capslash}[1]{{#1}\mskip -12mu / \mskip +4 mu}

\newcommand{\msp}{\;\;\;\;\;}
\newcommand{\boldm}[1]{\mbox{\boldmath ${#1}$}}

\begin{document}

%%%%%%%%%%%%%%%%%%%%%%%%%%%%%%%%%%%%%%%%%%%%%%%%%%%%%%%%%%%%%%%%%
%%%%%%%%%%%%%%%%%%%%%%%%%%%%%%%%%%%%%%%%%%%%%%%%%%%%%%%%%%%%%%%%%

\begin{flushright}
  HD-THEP-04-14\\
  hep-ph/0404006
\end{flushright}

\vspace{\baselineskip}

\begin{center}
\textbf{\LARGE Anomalous gauge-boson couplings\\[0.3em]
        and the Higgs-boson mass\\}
\vspace{4\baselineskip}
{\sc O.~Nachtmann\footnote{email:
O.Nachtmann@thphys.uni-heidelberg.de}, F.~Nagel\footnote{email:
F.Nagel@thphys.uni-heidelberg.de} and M.~Pospischil\footnote{Now at
CNRS UPR 2191, 1 Avenue de la Terrasse, F-91198 Gif-sur-Yvette,
France,\\\hspace*{.64cm}email: Martin.Pospischil@iaf.cnrs-gif.fr}}\\
\vspace{1\baselineskip}
\textit{Institut f\"ur Theoretische Physik, Philosophenweg 16, D-69120
Heidelberg, Germany}\\
\vspace{2\baselineskip}
\textbf{Abstract}\\
\vspace{1\baselineskip}
\parbox{0.9\textwidth}{We study anomalous gauge-boson couplings induced by a locally
\mbox{$SU(2)\times U(1)$} invariant effective Lagrangian containing ten
operators of dimension six built from boson fields of the Standard Model (SM)
before spontaneous symmetry breaking~(SSB).  After SSB some operators lead to
new three- and four-gauge-boson interactions, some contribute to the diagonal
and off-diagonal kinetic terms of the gauge bosons, to the kinetic term of the
Higgs boson and to the mass terms of the $W$ and $Z$~bosons.  This requires a
renormalisation of the gauge-boson fields, which, in turn, modifies the
charged- and neutral-current interactions, although none of the additional
operators contain fermion fields.  Also the Higgs field must be renormalised.
Bounds on the anomalous couplings from electroweak precision measurements at
LEP and SLC are correlated with the Higgs-boson mass~$m_H$.  Rather moderate
values of anomalous couplings allow $m_H$ up to 500~GeV.  At a future linear
collider the triple-gauge-boson couplings \mbox{$\gamma WW$} and \mbox{$ZWW$}
can be measured in the reaction \mbox{$e^+e^- \rightarrow WW$}.  We compare
three approaches to anomalous gauge-boson couplings: the form-factor approach,
the addition of anomalous coupling terms to the SM Lagrangian after and, as
outlined above, before~SSB.  The translation of the bounds on the couplings
from one approach to another is not straightforward.  We show that it can be
done for the process \mbox{$e^+e^- \rightarrow WW$} by defining new {\em
effective} \mbox{$\gamma WW$} and \mbox{$ZWW$} couplings.}
\end{center}
\vspace{\baselineskip}

\pagebreak

\tableofcontents

\pagebreak

%%%%%%%%%%%%%%%%%%%%%%%%%%%%%%%%%%%%%%%%%%%%%%%%%%%%%%%%%%%%%%%%%
%%%%%%%%%%%%%%%%%%%%%%%%%%%%%%%%%%%%%%%%%%%%%%%%%%%%%%%%%%%%%%%%%

\section{Introduction}
\label{sec-intro}

The Standard Model (SM) of particle physics has been tested in numerous
aspects with impressive success.  However, it lacks the attributes of a truly
fundamental theory since it does not predict the number of particles or
families and contains a large number of free parameters.  Moreover, it does
not incorporate gravity so that ultimately a different theory has to replace
the~SM.  One possibility is that physics beyond the SM will appear at an
energy scale~$\Lambda$.  From current electroweak precision fits one estimates
(see for instance~\cite{Kilian:2003yw}) that $\Lambda$ should be at least of
the order of TeV but, in fact, could be even much higher.  The impact of this
new high-scale physics on the phenomenology at lower energies can be taken
into account in various ways.

In the form-factor~(FF) approach the relevant vertices are parameterised in a
general way.  For the reaction \mbox{$e^+ e^- \rightarrow WW$} this was done
in~\cite{Gaemers:1978hg,Hagiwara:1986vm} for the three-gauge-boson vertices
\mbox{$\gamma WW$} and~\mbox{$ZWW$}.  There the structure of these two
vertices is only restricted by Lorentz invariance.  Form factors can and
should have imaginary parts.  Anomalous contributions to the \mbox{$\gamma
WW$}- and~\mbox{$ZWW$}-form factors have been studied extensively both for
LEP2 energies (see~\cite{Gounaris:1996rz} and references therein) and for the
energy range of future linear
colliders~\cite{Diehl:1993br,Diehl:1997ft,Menges:2001gg,Abe:2001wn,Bozovic-Jelisavcic:2002ta,Diehl:2002nj,Diehl:2003qz}.

Another possibility is to use an effective Lagrangian.  Here we have two
options.  We can start from the SM Lagrangian {\em after} spontaneous symmetry
breaking~(SSB) and add terms of higher dimension to obtain an effective
Lagrangian, which we call ELa~approach ({\em E}ffective {\em L}agrangian {\em
a}fter~SSB).  Alternatively we can start from the SM Lagrangian {\em before}
SSB and add terms of higher dimension there, called ELb~approach ({\em
E}ffective {\em L}agrangian {\em b}efore SSB).  In both cases the anomalous
coupling constants in the effective Lagrangian must be real.  Anomalous
imaginary parts in form factors are generated by loop effects using the
effective-Lagrangian techniques familiar from chiral perturbation theory, see
for instance~\cite{Weinberg:mt}.  The three approaches FF, ELa and ELb are
related but should not be confused with each other, see the discussion
in~\cite{Bernreuther:1996dh}.  The ELa approach, taking the anomalous terms in
leading order, produces only real parts of anomalous form factors.  In the ELb
approach the SSB has to be performed for the SM and the anomalous parts of the
Lagrangian together.  This has drastic consequences for all parts of the
Lagrangian as we shall analyse in detail in this paper for various electroweak
precision observables measured at LEP and SLC as well as for the reaction
\mbox{$e^+ e^- \rightarrow WW$} at a future linear
\mbox{$e^+e^-$}~collider~(LC).  It also has the consequence that the counting
of dimensions of anomalous terms is changed when Higgs fields are replaced by
their vacuum expectation values, see~\cite{Bernreuther:1996dh} where also the
question of \mbox{$SU(2) \times U(1)$} gauge invariance is discussed.
Anomalous couplings from operators of dimension $n$ in the ELb approach will
generate operators of dimension \mbox{$n' \leq n$} in the ELa approach.

Some advantages and disadvantages of the three approaches are as follows.  The
FF approach is the most general one but it has the disadvantage of introducing
many parameters.  Also, the anomalous parts of form factors for different
reactions like \mbox{$e^+ e^- \rightarrow WW$} and \mbox{$\gamma \gamma
\rightarrow WW$}  are {\em a priori} not related.  The ELa and ELb approaches
allow to relate anomalous effects in different reactions.  Suppose now that we
restrict the anomalous coupling terms to dimension \mbox{$n' \leq 6$} and
\mbox{$n \leq 6$} in the ELa and ELb approaches, respectively.  Then the ELa
approach generates more couplings than the ELb approach.  Thus, in a sense,
the ELb approach is the most restrictive framework if the dimension of the
coupling terms is limited.  For an application of the FF approach to the
reaction \mbox{$e^+ e^- \rightarrow \tau^+ \tau^-$} see for
instance~\cite{Bernreuther:1989kc}, for an application of the ELa approach to
$Z$ decays see~\cite{Bernreuther:jr}.  In the present paper we study mainly
the ELb approach to anomalous electroweak gauge-boson couplings.  We add to
the SM Lagrangian---before SSB---operators of higher dimension that consist of
SM fields.  The natural expansion parameter for this series is~$(v/\Lambda)$,
where \mbox{$v \approx 246$}~GeV is the vacuum expectation value of the
SM-Higgs-boson field.  Lists of all operators up to dimension six that respect
the SM gauge symmetry $SU(3) \times SU(2) \times U(1)$ were given
in~\cite{Buchmuller:1985jz,Leung:1984ni}, see also references therein.  A
number of studies of the effects of these operators for phenomenology were
made, see for instance~\cite{Hagiwara:1992eh,Hagiwara:1993ck}.  We will
comment below on the relation of these works to our present work.  Here we
follow~\cite{Buchmuller:1985jz} where systematic use is made of the equations
of motion in order to reduce the number of operators to an independent set.  A
particularly interesting part of this Lagrangian is its gauge-boson sector
because, in the~SM, the structure of the gauge-boson vertices is highly
restricted.  In the SM there exist triple- as well as quartic-gauge-boson
couplings all of which are fixed by the coupling constants of $SU(2)$ and
$U(1)$, see for instance~\cite{Nachtmann:ta}.  At tree-level the triple
couplings \mbox{$\gamma WW$}, \mbox{$ZWW$} and only the quartic couplings
\mbox{$WWWW$}, \mbox{$\gamma \gamma WW$}, \mbox{$\gamma ZWW$} and
\mbox{$ZZWW$} occur.  Furthermore, in the SM the interactions of gauge bosons
with the Higgs boson are determined by the covariant derivative acting on the
Higgs field.

Here we consider the leading order operators of dimension higher than
four---that is of dimension six---that consist either only of electroweak
gauge-boson fields or of gauge-boson fields combined with the Higgs-boson
field of the~SM.  There are ten such operators, four of them $CP$~violating,
see~\cite{Buchmuller:1985jz}.  This leads to ten new coupling constants~$h_i$,
subsequently called anomalous couplings, which parameterise deviations from
the~SM.  It is assumed that the new-physics scale $\Lambda$ is large enough
such that operators of dimension six already give a good description of the
high-scale effects.  To keep the number of anomalous couplings within
reasonable limits we exclude all non-SM operators that {\em a priori} involve
fermions.  Nevertheless, the purely bosonic anomalous couplings change the
gauge-boson-fermion interactions in the following way.  After SSB the pure
boson operators contribute to the diagonal as well as off-diagonal kinetic
terms of the gauge bosons and to the mass terms of the $W$ and $Z$ bosons.
Firstly, this requires a renormalisation of the $W$-boson field.  Secondly,
the kinetic and the mass matrices of the neutral gauge bosons have to be
diagonalised simultaneously to obtain the physical photon and $Z$-boson fields
as linear combinations of the photon and $Z$-boson fields of the effective
Lagrangian.  This in turn modifies the neutral- and charged-current
interactions.  Since all fermion families are affected in the same manner no
flavour-changing neutral currents are induced.  Moreover two dimension-six
operators contribute to the kinetic term of the Higgs boson such that a
renormalisation of the Higgs field is necessary, too.

Thus in the ELb approach purely bosonic anomalous couplings influence also the
precision observables from $Z$~decay.  In this paper we exploit this to
calculate bounds on two $CP$~conserving anomalous couplings from measurements
at LEP1 and SLC and from $W$-boson measurements.  To this end precision
observables that are sensitive to the modified gauge-boson-fermion
interactions or to the mass of the $W$~boson are used.  Less stringent bounds
are obtained from direct measurements of the three-gauge-boson vertices
\mbox{$\gamma WW$} and \mbox{$ZWW$} in various processes at LEP2.  However,
one more $CP$~conserving coupling and two $CP$~violating couplings can be
constrained using this data.  Bounds on anomalous triple-gauge-boson couplings
(TGCs) have been measured by the CDF~collaboration~\cite{Abe:1994fx} and the
\mbox{DO\hspace{-.23cm}\slash}\hspace{.15cm}collaboration~\cite{Abbott:1999ec}
and are discussed in Sect.~6.1.

One important purpose of future high-energy experiments is the precision check
of the relations between the various gauge-boson couplings.  Their SM values
guarantee the renormalisability of the electroweak theory.  Thus any observed
deviations from these SM values would have drastic consequences for the
structure of the theory.  Gauge-boson couplings can be studied at the
LHC~\cite{Kuss:1997mf,Green:2003mn,Diakonos:1992qc} and with high precision at
a future LC like
TESLA~\cite{Richard:2001qm,Aguilar-Saavedra:2001rg,Monig:2003cm},
NLC~\cite{Abe:2001wn}, JLC~\cite{Abe:2001gc} or CLIC~\cite{Ellis:1998wx}.
There $W$~pair production, \mbox{$e^+ e^- \rightarrow WW$}, is suitable to
measure~TGCs.  In previous
work~\cite{Diehl:1993br,Diehl:1997ft,Diehl:2002nj,Diehl:2003qz} on \mbox{$e^+
e^- \rightarrow WW$} by our group we followed the form-factor (FF) approach
using the parameterisation of the \mbox{$\gamma WW$} and \mbox{$ZWW$} vertices
of~\cite{Hagiwara:1986vm}.  The maximum achievable sensitivity to the
anomalous couplings in this process at c.m.~energies of 500~GeV, 800~GeV and
3~TeV was determined by means of optimal
observables~\cite{Diehl:1993br,Diehl:1997ft} for the case of no or
longitudinal beam polarisation in~\cite{Diehl:2002nj}, and for transverse beam
polarisation in~\cite{Diehl:2003qz}.  Optimal observables were introduced for
one-variable problems in~\cite{Atwood:1991ka} and for multi-variable problems
in~\cite{Diehl:1993br}.  In the present paper we use, as explained above, the
effective Lagrangian approach~ELb.  We give a detailed comparison of the FF
and the ELb approaches for \mbox{$e^+ e^- \rightarrow WW$} in the following.
In our ELb approach not only the \mbox{$\gamma WW$} and \mbox{$ZWW$}~vertices
but also the gauge-boson-fermion vertices and the $W$ and $Z$~propagators get
anomalous contributions.  We show that nevertheless the results computed in
the FF approach can be transformed into bounds on the anomalous couplings used
here with~ELb.  This is achieved be defining new {\it effective} \mbox{$\gamma
WW$} and \mbox{$ZWW$} couplings that are specific for the reaction~\mbox{$e^+
e^- \rightarrow WW$}.  In our ELb~approach we have \mbox{$SU(2) \times U(1)$}
gauge invariance and we have restricted ourselves to dimension six for the
additional operators.  These two ingredients together lead to the well known
``gauge relations'' for the TGCs~\cite{Gounaris:1996rz}.  Note that
\mbox{$SU(2) \times U(1)$} gauge invariance alone gives no restrictions on
the~TGCs.  Interestingly we find that even if the usual gauge relations hold
for the original TGCs these relations change when we use the effective
couplings, which are directly related to the FF approach.  Moreover, even
without {\em effective} couplings the shape of the gauge relations depends on
the input parameter scheme.

In this paper we also mention some properties of the \mbox{$\gamma \gamma WW$}
and \mbox{$\gamma \gamma H$}~vertices that do not occur in the observables
that we consider here but play an important r\^ole in the reaction
\mbox{$\gamma \gamma \rightarrow WW$} at a collider with two high-energy
photons in the initial state.  Such a photon collider has been proposed as an
option for TESLA~\cite{Badelek:2001xb} and for CLIC~\cite{Burkhardt:2002vh}.
The process \mbox{$\gamma \gamma \rightarrow WW$} will be studied in
forthcoming work~\cite{photon-collider}.  Clearly, for a comparison of the
reactions \mbox{$e^+ e^- \rightarrow WW$} and \mbox{$\gamma \gamma \rightarrow
WW$} the ELb framework is the most suitable one.  This is the main motivation
for treating \mbox{$e^+ e^- \rightarrow WW$} in the ELb approach in the
present paper, since our results here are required for the discussion of
\mbox{$\gamma \gamma \rightarrow WW$} in~\cite{photon-collider}.  There we
shall give a comparison of the sensitivities of the reactions \mbox{$e^+ e^-
\rightarrow WW$} and \mbox{$\gamma \gamma \rightarrow WW$} to anomalous
gauge-boson couplings.

This work is organised as follows: In Sect.~\ref{sec-eff} we give an overview
of the operators in our effective Lagrangian (ELb approach) and explain, which
operators contribute to the kinetic and mass terms of the gauge bosons and of
the Higgs boson, to the three- and four-gauge-boson couplings, and to the
photon-photon-Higgs coupling.  In Sect.~\ref{sec-diag} we perform the
simultaneous diagonalisation of the kinetic and mass terms of the neutral
gauge bosons and the renormalisation of the charged gauge boson and Higgs
boson fields.  We then consider the interactions of gauge bosons with fermions
in~Sect.~\ref{sec-para} and define two different sets of electroweak
parameters, that we use to calculate the observables: one set, $P_Z$,
containing the $Z$~mass, the other one, $P_W$, containing the $W$~mass.  In
Sect.~\ref{sec-lep} we present the bounds on the anomalous couplings from
electroweak precision measurements at LEP and SLC, except for direct
measurements of the three-gauge-boson vertices, thereby using $P_Z$.  In
Sect.~\ref{sec-tgcs} we give the relations of the standard couplings
\mbox{$\Delta g_1^{\gamma}$}, \mbox{$\Delta \kappa_{\gamma}$}, etc.\ for the
\mbox{$\gamma WW$}~and \mbox{$ZWW$}~vertices to our anomalous couplings using
$P_Z$ and, alternatively, using $P_W$ as input parameters.  We derive bounds
on the anomalous couplings of the effective Lagrangian from measurements of
TGCs at LEP2 using~$P_Z$.  We analyse in detail the reaction \mbox{$e^+e^-
\rightarrow WW$} at a future LC where we define effective \mbox{$\gamma
WW$}~and \mbox{$ZWW$}~couplings using $P_W$.  We calculate the bounds
obtainable on the anomalous couplings using the results
of~\cite{Diehl:2002nj,Diehl:2003qz} for this reaction.  In
Sect.~\ref{sec-conc} we present our conclusions.

%%%%%%%%%%%%%%%%%%%%%%%%%%%%%%%%%%%%%%%%%%%%%%%%%%%%%%%%%%%%%%%%%
%%%%%%%%%%%%%%%%%%%%%%%%%%%%%%%%%%%%%%%%%%%%%%%%%%%%%%%%%%%%%%%%%

\section{Effective Lagrangian}
\label{sec-eff}
\setcounter{equation}{0}

Our starting point is the effective Lagrange density $\mathscr{L}_{\rm eff}$
containing all lepton- and baryon-number-conserving operators that can be
built from SM fields, see~\cite{Buchmuller:1985jz}.  Let $\Lambda$ be the
scale of new physics and \mbox{$v \approx 246$}~GeV be the vacuum expectation
value of the Higgs field.  If not stated otherwise, numerical values of
physical parameters are taken from~\cite{Hagiwara:pw}.  Throughout this paper
we assume
\be
\Lambda \gg v.
\ee
Then $\mathscr{L}_{\rm eff}$ can be expanded as 
\be
\label{eq-expans}
\mathscr{L}_{\rm eff} = \mathscr{L}_{0} + \mathscr{L}_{1} + \mathscr{L}_{2} +
\ldots \;,
\ee
where $\mathscr{L}_{0}$ contains operators of dimension less or equal to four,
$\mathscr{L}_{1}$ of dimension five, $\mathscr{L}_{2}$ of dimension six etc.
The terms $\mathscr{L}_{1}$, $\mathscr{L}_{2}, \ldots$ give contributions of
order \mbox{$(v/\Lambda)$}, \mbox{$(v/\Lambda)^2,\ldots$} in the amplitudes,
thus~(\ref{eq-expans}) represents effectively an expansion in powers
of~$(v/\Lambda)$.

Given the SM particle content, the general form of $\mathscr{L}_{0}$ is fixed
as that of the SM Lagrangian by gauge invariance.  For the SM Lagrangian we
use the conventions of~\cite{Nachtmann:ta}.  Restricting ourselves to the
electroweak interactions and neglecting neutrino masses we have (see Chapter
22 of~\cite{Nachtmann:ta})
\bea
\label{eq:smlagr}
\mathscr{L}_0 & = &
- \frac{1}{4} W^i_{\mu \nu} W^{i \, \mu \nu} 
- \frac{1}{4} B_{\mu \nu} B^{\mu \nu}\\
& & \mbox{} + \left(\mathcal{D}_{\mu} \varphi \right)^{\dag}
\left(\mathcal{D}^{\mu} \varphi\right)
+ \mu^2 \varphi^{\dag} \varphi - \lambda \left(
\varphi^{\dag} \varphi \right)^2 \nonumber\\
& & \mbox{} + i \overline{L}  \capslash{\mathcal{D}} L 
+ i \overline{E} \capslash{\mathcal{D}} E
+ i \overline{Q} \capslash{\mathcal{D}} Q 
+ i \overline{U} \capslash{\mathcal{D}} U 
+ i \overline{D} \capslash{\mathcal{D}} D \nonumber \\
& & \mbox{} - \left(\overline{E}\, \Gamma_E \, \varphi^{\dag} L 
+ \overline{U}\, \Gamma_U \, \tilde{\varphi}^{\dag} Q
+ \overline{D}\, \Gamma_D \, \varphi^{\dag} Q + {\rm H.c.}
\right).\nonumber
\eea
The \mbox{$3\times 3$}-Yukawa matrices have the form
\bea
\label{eq-gammae}
\Gamma_E & = & {\rm diag} (c_{\rm e}, c_{\mu}, c_{\tau}),\\
\label{eq-gammau}
\Gamma_U & = & {\rm diag} (c_{\rm u}, c_{\rm c}, c_{\rm t}),\\
\label{eq-gammad}
\Gamma_D & = & V {\rm diag} (c_{\rm d}, c_{\rm s}, c_{\rm b}) V^{\dag},
\eea
where the diagonal elements all obey \mbox{$c_i \geq 0$} and $V$ is the
CKM~matrix.  With these conventions the matrices $\Gamma_E$, $\Gamma_U$,
$\Gamma_D$ correspond to the matrices $C_{\ell}$, $C'_q$, $C_q$
in~\cite{Nachtmann:ta}, respectively.  The vector of the three left-handed
lepton doublets is denoted by $L$, of the right-handed charged leptons by $E$,
of the left-handed quark doublets by $Q$, and of the right-handed up- and
down-type quarks by $U$ and~$D$.  The Higgs field is denoted by~$\varphi$ and
we define
\be
\tilde{\varphi} = \varepsilon \varphi^*,\msp \msp
\varepsilon = \left( \ba{rr} 0 & 1 \\ -1 & 0 \ea \right).
\ee
The covariant derivative is
\be
\label{eq-covderiva}
\mathcal{D}_{\mu} = \partial_{\mu} + i g  W^i_{\mu} \mathbf{T}_i + i g' B_{\mu}
\mathbf{Y},
\ee
where $\mathbf{T}_i$ and $\mathbf{Y}$ are the generating operators of
weak-isospin and weak-hypercharge transformations.  For the left-handed
fermion fields and the Higgs doublet we have \mbox{$\mathbf{T}_i = \tau_i /
2$}, where $\tau_i$ are the Pauli matrices.  For the right-handed fermion
fields we have \mbox{$\mathbf{T}_i = 0$}.  The hypercharges $y$ of the
fermions and the Higgs doublet are listed in Tab.~\ref{tab:hyper}.
\begin{table}
\centering
\begin{tabular}{cccccccc}
\hline
 &&&&&&&\\[-.32cm]
 & $L$ & $E$ & $Q$ & $U$ & $D$ & $\varphi$\\
\hline
 &&&&&&&\\[-.25cm]
$y$ & $-\frac{1}{2}$ & $-1$ & $\frac{1}{6}$ & $\frac{2}{3}$ & $-\frac{1}{3}$ &
$\frac{1}{2}$\\[.1cm]
\hline
\end{tabular}
\caption{\label{tab:hyper}Weak hypercharge of the fermions and the Higgs
 doublet.}
\end{table}
The field strengths are\footnote{The signs in front of the gauge couplings
in~(\protect{\ref{eq-covderiva}}) and~(\protect{\ref{eq-gifs}}) differ from
the conventions of~\protect{\cite{Buchmuller:1985jz}}.  This may lead to sign
changes in the dimension-six operators discussed below.}
\bea
\label{eq-gifs}
W^i_{\mu \nu}  & = & \partial_{\mu}W^i_{\nu} - \partial_{\nu}W^i_{\mu}
- g \, \epsilon_{ijk} \, W^j_{\mu} W^k_{\nu},\\
B_{\mu \nu} & = & \partial_{\mu}B_{\nu} - \partial_{\nu}B_{\mu}.\nonumber
\eea
For the parameters of the Higgs potential in~(\ref{eq:smlagr}) we assume
\be
\mu^2 > 0,\msp \msp \lambda > 0.
\ee
Then the potential has a minimum for constant field satisfying
\be
\label{eq-vdef}
\sqrt{2 \varphi^{\dagger} \varphi} = \sqrt{\frac{\mu^2}{\lambda}} \equiv v.
\ee
After SSB, that is in the unitary gauge, we can choose the Higgs field to have
the form
\be
\label{eq-Higgs}
\varphi(x) = \frac{1}{\sqrt{2}} \left(\ba{c}0 \\ v + H'(x)\ea\right),
\ee
where $H'(x)$ would be the physical Higgs field in the~SM, and in lowest order
the vacuum expectation value of the Higgs field, $v$, is given in terms of the
Lagrangian parameters by~(\ref{eq-vdef}).  Looking at the Higgs-mass term we
find for the squared mass of the Higgs boson in the~SM
\be
\label{eq-Higgsmass}
m_H^{\prime \, 2} = 2 \lambda v^2.
\ee
The coupling constants in~(\ref{eq-gammae}) to~(\ref{eq-gammad}) are related
to the fermion masses by
\be
m_{\rm j} = c_{\rm j} \frac{v}{\sqrt{2}}
\ee
with \mbox{${\rm j} = {\rm u},\, {\rm c},\, {\rm t},\, {\rm d},\, {\rm s},\,
  {\rm b}, {\rm e},\, {\mu},\, {\tau}$}.

The higher-dimensional operators in $\mathscr{L}_{1}$, $\mathscr{L}_{2}$ etc.\
in~(\ref{eq-expans}) describe the effects of new physics at the scale
$\Lambda$ on the phenomenology at the weak scale~$v$.
Following~\cite{Buchmuller:1985jz,Leung:1984ni}, we assume \mbox{$SU(3) \times
SU(2) \times U(1)$} gauge invariance also for the new interactions.  The only
Lorentz and gauge invariant operator of dimension five that can be constructed
from SM fields violates lepton-number conservation, see
\cite{Buchmuller:1985jz}, and hence is not considered here.  Thus, the
leading-order addition to the SM Lagrangian is $\mathscr{L}_{2}$, which should
therefore lead to a good description of the new-physics effects at energies
sufficiently below~$\Lambda$.  Compared to~\cite{Leung:1984ni} the number of
operators of dimension six to be considered is reduced
in~\cite{Buchmuller:1985jz} by systematically applying the equations of
motion.  This is a completely legitimate procedure for our purposes, see also
the discussion of this point in~\cite{Bernreuther:1996dh}.  We thus refer to
the list of operators in~\cite{Buchmuller:1985jz} for our analysis.

Out of the 80 dimension-six operators listed in~\cite{Buchmuller:1985jz} we
consider all operators that consist either only of electroweak gauge-boson
fields or of gauge-boson fields combined with the SM Higgs field, see
\mbox{$(3.5)$}, \mbox{$(3.6)$} and \mbox{$(3.41)$} to \mbox{$(3.44)$}
in~\cite{Buchmuller:1985jz}:
\begin{alignat}{4}
\label{eq-Ow}
&O_W && \! = \epsilon_{ijk} \, W^{i \, \nu}_{\mu} W^{j \, \lambda}_{\nu}
W^{k \, \mu}_{\lambda},\phantom{\frac{1}{2}}&\!\!\!\!\!
&O_{\tilde{W}} && \! = \epsilon_{ijk} \, \tilde{W}^{i \, \nu} _{\mu} W^{j
\, \lambda} _{\nu} W^{k \, \mu}_{\lambda},\\[.1cm]
\label{eq-Ophiw}
&O_{\varphi W} && \! = \frac{1}{2} \left( \varphi^{\dag}\varphi
\right)\,W^i_{\mu \nu} W^{i\, \mu \nu},&\!\!\!\!\!
&O_{\varphi \tilde{W}} && \! = \left( \varphi^{\dag} \varphi
\right)\,\tilde{W}^i_{\mu \nu} W^{i \, \mu \nu},\\[.1cm]
\label{eq-Ophib}
&O_{\varphi B} && \! = \frac{1}{2} \left( \varphi^{\dag}\varphi
\right)\,B_{\mu \nu} B^{\mu \nu},&\!\!\!\!\!
&O_{\varphi \tilde{B}} && \! = \left( \varphi^{\dag}\varphi
\right)\,\tilde{B}_{\mu \nu} B^{\mu \nu},\\[.1cm]
\label{eq-Owb}
&O_{W \mskip -3mu B} && \! = \left( \varphi^{\dag} \tau^i \varphi
\right)\,W^i_{\mu \nu}B^{\mu \nu},\phantom{\frac{1}{2}}&\!\!\!\!\!
&O_{\tilde{W} \mskip -3mu B} && \! = \left( \varphi^{\dag} \tau^i \varphi
\right)\,\tilde{W}^i_{\mu \nu} B^{\mu \nu},\\[.1cm]
\label{eq-Ophi1}
&O_{\varphi}^{(1)} && \!\!\!\! = \left( \varphi^{\dag}\varphi \right) \left(
\mathcal{D}_{\mu} \varphi \right)^{\dag} \left(\mathcal{D}^{\mu} \varphi
\right),\phantom{\frac{1}{2}}&\!\!\!\!\!
&O_{\varphi}^{(3)} && \!\!\!\! = \left(\varphi^{\dag} \mathcal{D}_{\mu} \varphi
\right)^{\dagger}\left( \varphi^{\dag} \mathcal{D}^{\mu} \varphi \right).
\end{alignat}
Here the dual field strengths are defined as
\be
\tilde{W}^i_{\mu \nu} = \frac{1}{2} \epsilon_{\mu\nu\rho\sigma}
W^{i\,\rho\sigma},\msp \msp
\tilde{B}_{\mu \nu} = \frac{1}{2} \epsilon_{\mu\nu\rho\sigma}
B^{\rho\sigma}.
\ee

In the following we therefore use the effective Lagrangian
\be
\label{eq-Leff}
\mathscr{L}_{\rm eff} = \mathscr{L}_{0} + \mathscr{L}_{2},
\ee
where $\mathscr{L}_{0}$ is the SM part~(\ref{eq:smlagr}).  The non-SM part
with the dimension-six operators is
\bea
\mathscr{L}_{2} & = &
\Big(h_W O_W +
h_{\tilde{W}} O_{\tilde{W}}
 + h_{\varphi W} O_{\varphi W} +
h_{\varphi \tilde{W}} O_{\varphi \tilde{W}}\nonumber\\
 & & \mbox{} + h_{\varphi B} O_{\varphi B} +
h_{\varphi \tilde{B}} O_{\varphi \tilde{B}}
 + h_{W \mskip -3mu B} O_{W \mskip -3mu B} +
h_{\tilde{W} \mskip -3mu B} O_{\tilde{W} \mskip -3mu B}\nonumber\\
 & & \mbox{} + h_{\varphi}^{(1)} O_{\varphi}^{(1)} +
h_{\varphi}^{(3)} O_{\varphi}^{(3)}\Big)/v^2,
\label{eq-Leff2}
\eea
where we have divided by $v^2$ in order to obtain dimensionless coupling
constants $h_i$, with \mbox{$i = W, \tilde{W}, \varphi W,\ldots$}.  The~$h_i$
are subsequently called anomalous couplings.  Nominally we have
\be
\label{eq-order}
h_i = O(v^2/\Lambda^2).
\ee
%
%

%%%%%%%%%%%%%%%%%%%%%%%%%%%%%%%%%%%%%%%%%%%%%%%%%%%%%%%%%%%%%%%%%
%%%%%%%%%%%%%%%%%%%%%%%%%%%%%%%%%%%%%%%%%%%%%%%%%%%%%%%%%%%%%%%%%

\section{Symmetry breaking and diagonalisation in the gauge-boson sector}
\label{sec-diag}
\setcounter{equation}{0}

Starting from the Lagrangian~(\ref{eq-Leff}) we go now to the unitary gauge,
that is we replace the Higgs field everywhere by the
expression~(\ref{eq-Higgs}) involving only the Higgs-vacuum-expectation value
$v$ and the field~\mbox{$H'(x)$}, which would be the physical Higgs-boson
field for zero anomalous couplings.  If this is done for $\mathscr{L}_{0}$ we
arrive at the SM~Lagrangian in unitary gauge, see~(22.123)
of~\cite{Nachtmann:ta}.  It is convenient to take this as starting point and
consider the necessary changes due to the $\mathscr{L}_{2}$ term
in~(\ref{eq-Leff}) subsequently.  Let us, therefore, introduce boson fields
$A_{\mu}^{\prime}$, $Z_{\mu}^{\prime}$ and $W_{\mu}^{\prime \pm}$ which would
be the physical gauge-boson fields if we considered only the SM
Lagrangian~$\mathscr{L}_{0}$.  The original $W_{\mu}^i$ and $B_{\mu}$ fields
are expressed in terms of these fields as follows:
\bea
\label{eq-prfielddef1}
\!\!\!\!\!\!\!\! W_{\mu}^1 & = & \frac{1}{\sqrt{2}} \left(W_{\mu}^{\prime +} +
W_{\mu}^{\prime -}\right),
\qquad \!\!\!\!\!\!\!\!\!\! W^3_{\mu} = c_{\rm w}' \, Z'_{\mu} + s_{\rm w}' \,
A'_{\mu}\;,\\
\label{eq-prfielddef2}
\!\!\!\!\!\!\!\! W_{\mu}^2 & = & \frac{i}{\sqrt{2}} \left(W_{\mu}^{\prime +} -
W_{\mu}^{\prime -}\right),
\qquad \!\!\!\!\!\!\!\!\!\! B_{\mu} = - s_{\rm w}' \, Z'_{\mu} + c_{\rm w}' \,
A'_{\mu}\;,
\eea
where
\begin{alignat}{3}
\label{eq-sinepr}
&s'_{\rm w} \; & \equiv & \; \sin \theta'_{\rm w} \; & = & \;
      \frac{g'}{\sqrt{g^2 + g^{\prime \, 2}}},\\
\label{eq-cosinepr} 
&c'_{\rm w} \; & \equiv & \; \cos \theta'_{\rm w} \; & = & \;
      \frac{g}{\sqrt{g^2 + g^{\prime \, 2}}}
\end{alignat}
are the sine and cosine of the weak mixing angle in the SM, determined by the
\mbox{$SU(2)$} and \mbox{$U(1)_Y$} couplings of~$\mathscr{L}_{0}$.  Without
loss of generality we can assume $g$ and $g'$ to be greater than zero and
therefore have~\mbox{$0 \leq \theta'_{\rm w} \leq \pi/2$}.  The positron
charge $e'$ of~$\mathscr{L}_{0}$ is given by
\be
\label{eq-epr}
e' = g s'_{\rm w}.
\ee
The next step is to consider the term~$\mathscr{L}_2$ in~(\ref{eq-Leff}),
(\ref{eq-Leff2}), and insert for the Higgs field \mbox{$\varphi (x)$}
everywhere (\ref{eq-Higgs}) and for the gauge-boson fields
(\ref{eq-prfielddef1}),~(\ref{eq-prfielddef2}).  We see then easily that the
original dimension-six operators in~$\mathscr{L}_2$ give now contributions to
dimension-two, -three, -four, -five and -six terms.

In Tab.~\ref{tab:contri} we list from which coupling constants
in~(\ref{eq-Leff2}) corresponding to the operators (\ref{eq-Ow}) to
(\ref{eq-Ophi1}) we get contributions to the kinetic and mass terms of the
gauge bosons, to the kinetic terms of the Higgs boson, and to several coupling
terms of the gauge bosons and the Higgs boson in the basis $W^{\prime \pm}$,
$Z'$, $A'$,~$H'$.  The kinetic terms of the gauge bosons receive contributions
only from $O_{\varphi W}$, $O_{\varphi B}$ and $O_{W \mskip -3mu B}$.  The
operators $O_{\varphi \tilde{W}}$, $O_{\varphi \tilde{B}}$ or $O_{\tilde{W}
\mskip -3mu B}$ do not contribute there since their terms of second order in
the boson fields vanish after partial integration.  The operators
$O_{\varphi}^{(1)}$ and $O_{\varphi}^{(3)}$ contribute only to the
gauge-boson-mass terms and to the kinetic term of the Higgs field~$H'$.

In Tab.~\ref{tab:contri} we also show how the dimension-six operators
contribute to those gauge-boson and gauge-boson-Higgs vertices that are
required for our studies.
\begin{table*}
\centering
\begin{tabular}{rccccccccccc}
\hline
 &&&&&&&&&&&\\[-.28cm]
 & SM & $h_W$ & $h_{\tilde{W}}$ & $h_{\varphi
 W}$ & $h_{\varphi \tilde{W}}$ & $h_{\varphi B}$ &
 $h_{\varphi \tilde{B}}$ & $h_{W \mskip -3mu B}$ &
 $h_{\tilde{W} \mskip -3mu B}$ &
 $h_{\varphi}^{(1)}$ & $h_{\varphi}^{(3)}$\\[0.06cm]
\hline
 &&&&&&&&&&&\\[-.25cm]
gauge kinetic & $\surd$ & & & $\surd$ & & $\surd$ & & $\surd$ & & & \\
gauge mass & $\surd$ & & & & & & & & &  $\surd$ & $\surd$ \\
Higgs kinetic & $\surd$ & &&&&&&&& $\surd$ & $\surd$ \\
$V'W^{\prime +}W^{\prime -}$ & $\surd$ & $\surd$ & $\surd$ & $\surd$ & & & &
 $\surd$ & $\surd$ & &\\
$A'A'W^{\prime +}W^{\prime -}$ & $\surd$ & $\surd$ & $\surd$ & $\surd$ & & & &
 & & & \\ 
$A'A'H'$ & & & & $\surd$ & $\surd$ & $\surd$ & $\surd$ & $\surd$ &
 $\surd$ & \\[.05cm]
\hline
\end{tabular}
\caption{\label{tab:contri}Contributions from SM Lagrangian and from
operators (\ref{eq-Ow}) to (\ref{eq-Ophi1}) to kinetic and mass terms of gauge
bosons, to the kinetic term of the Higgs boson and to terms of the form
\mbox{$V'W^{\prime +}W^{\prime -}$}, \mbox{$A'A'W^{\prime +}W^{\prime -}$} and
\mbox{$A'A'H$} with \mbox{$V' = A'$} or $Z'$.  Note that the contributions to
the {\em physical} \mbox{$\gamma WW$}, \mbox{$ZWW$} and \mbox{$\gamma \gamma
H'$} vertices {\em after} the simultaneous diagonalisation are different, see
Tab.~\protect{\ref{tab:vertices}} below.}
\end{table*}
Note that in Tab.~\ref{tab:contri} we show the contributions to the vertices
where the operators are still written in terms of the primed fields $W^{\prime
\pm}$, $Z'$, $A'$,~$H'$.  The operators $O_W$ and $O_{\tilde{W}}$ contribute
both to the three- and to the four-gauge-boson couplings.  The operators
$O_{\varphi W}$, $O_{W \mskip -3mu B}$ and $O_{\tilde{W} B}$ contribute to the
three-gauge-boson vertices with terms proportional to~$v^2$.  In addition, the
operator $O_{\varphi W}$ also induces a four-gauge-boson vertex.  The operator
$O_{\varphi \tilde{W}}$ contributes neither to the TGCs, since the
corresponding term can be written as a total divergence, nor to the
four-gauge-boson couplings because the term of the form
\be
\epsilon^{\mu \nu \rho \sigma} \epsilon_{ijk} \epsilon_{ilm} W^j_{\mu}
W^k_{\nu} W^l_{\rho} W^m_{\sigma}
\ee
vanishes for symmetry reasons.  In addition, six operators give rise to a
\mbox{$A'A'H'$}~vertex.  The dimension-six operators of~$\mathscr{L}_{2}$
induce anomalous terms to further vertices, e.g.\ \mbox{$Z'Z'H'$} and
\mbox{$W^{\prime +}W^{\prime -}H'$}, which are however not relevant for our
calculations.

We see that with the inclusion of~$\mathscr{L}_{2}$, the kinetic and the mass
terms of the gauge bosons as well as the kinetic term of the Higgs field~$H'$
do not have standard form any more due to additional contributions arising
according to Tab.~\ref{tab:contri}.  We have now to diagonalise the mass
matrix and simultaneously transform the kinetic matrix to the unit matrix to
identify the physical gauge-boson fields and the physical Higgs-boson field.
The gauge-boson kinetic and mass terms of the effective
Lagrangian~(\ref{eq-Leff}) are given by
\be
\label{eq-quadlag}
\mathscr{L}^{(2)}_V+\mathscr{L}^{(2)}_W \;, 
\ee
where
\bea
\label{eq-vprime}
\mathscr{L}^{(2)}_V & = &-\frac{1}{4}\boldsymbol{V}_{\mu \nu}
^{\prime \, \rm T} \,\, T' \,\, \boldsymbol{V}'^{\mu \nu} + \frac{1}{2}
\boldsymbol{V}^{\prime \, \rm T}_{\mu} \,\, M' \,\,
\boldsymbol{V}'^{\mu}\;,\\[.2cm]
\label{eq-wquad}
\mathscr{L}^{(2)}_W & = &-\frac{1}{2} \left(1 - h_{\varphi W}\right) W^{\prime
  +}_{\mu \nu}W^{\prime - \mu \nu}\\
& & \mbox{} + m^{\prime\,2}_W \left(1 +
h_{\varphi}^{(1)}/2\right)  W^{\prime +}_{\mu}W^{\prime -
  \mu}\;,\nonumber\\[.2cm] 
\boldsymbol{V}'_{\mu \nu} & = & \partial_{\mu}\boldsymbol{V}'_{\nu}-
\partial_{\nu}\boldsymbol{V}'_{\mu}\;,\;\;\;\;
\boldsymbol{V}^{\prime}_{\mu} = \left(Z'_{\mu},A'_{\mu}\right)^{\rm T},\\[.2cm]
W_{\mu \nu}^{\prime \pm} & = & \partial_{\mu} W_{\nu}^{\prime \pm} -
\partial_{\nu} W_{\mu}^{\prime \pm}.
\eea
Here we have introduced vector notation for the neutral primed gauge fields,
and $T'$ and $M'$ are given by
\be
\label{eq-masskin}
T' = \left( \begin{array}{*{2}{c}} a & b  \\ b &
d \end{array}\right),\;\;\;\;\;\;
M' = m_Z^{\prime \,2} \Big(1 + \frac{1}{2} \left(h_{\varphi}^{(1)} +
h_{\varphi}^{(3)}\right) \Big) \left( \begin{array}{*{2}{c}} 1 & 0
\\ 0 & 0 \end{array} \right)
\ee
with
\bea
\label{eq-adef}
a & = & 1 - 2 c'_{\rm w} s'_{\rm w} h_{W \mskip -3mu B} - c_{\rm
w}^{\prime \, 2} h_{\varphi W} - s_{\rm w}^{\prime \, 2} h_{\varphi
B},\\[.2cm]
b & = & \left(c_{\rm w}^{\prime \, 2} - s_{\rm w}^{\prime \,
2}\right) h_{W \mskip -3mu B} + c'_{\rm w} s'_{\rm w}
\left(h_{\varphi B} - h_{\varphi W}\right),\\[.2cm]
\label{eq-ddef}
d & = & 1 + 2 c'_{\rm w} s'_{\rm w} h_{W \mskip -3mu B} - s_{\rm
w}^{\prime \, 2} h_{\varphi W} - c_{\rm w}^{\prime \, 2} h_{\varphi B}.  
\eea
The quantities 
\bea
\label{eq-prwmass}
m^{\prime\,2}_W & = & g^2 v^2/4,\\
\label{eq-przmass}
m_Z^{\prime \,2} & = & (g^2 + g^{\prime \,2}) v^2/4.
\eea
would be the squared gauge-boson masses after SSB if we considered only the SM
Lagrangian~$\mathscr{L}_{0}$.  Because of charge conservation there is no
mixing between charged and neutral gauge-boson fields in~(\ref{eq-quadlag}).
Moreover, the matrix~$M'$ has only one non-zero entry (corresponding to
\mbox{$Z'Z'$}) since terms of second order in the gauge fields without
derivatives can only come from operators with two covariant derivatives of
Higgs fields, as occuring in (\ref{eq:smlagr}) and~(\ref{eq-Ophi1}).  There, due
to~(\ref{eq-Higgs}), only the massive gauge bosons contribute.

We would like to find a basis in the fields such that~(\ref{eq-quadlag}) takes
the standard form:
\bea
\label{eq-lv}
\!\!\!\!\! \mathscr{L}^{(2)}_V & = & - \frac{1}{4} \left(Z_{\mu \nu} Z^{\mu
    \nu} + A_{\mu \nu} A^{\mu \nu}\right) + \frac{1}{2} m_Z^2 Z_{\mu}
    Z^{\mu},\\
\label{eq-lw}
\!\!\!\!\! \mathscr{L}^{(2)}_W & = & -\frac{1}{2}W^{+}_{\mu \nu}W^{- \mu \nu} +
m_W^2 W^{+}_{\mu}W^{- \mu},
\eea
where
\bea
Z_{\mu \nu} & = & \partial_{\mu} Z_{\nu} - \partial_{\nu} Z_{\mu},\\[.3em]
A_{\mu \nu} & = & \partial_{\mu} A_{\nu} - \partial_{\nu} A_{\mu},\\[.3em]
W_{\mu \nu}^{\pm} & = & \partial_{\mu} W_{\nu}^{\pm} - \partial_{\nu}
  W_{\mu}^{\pm},
\eea
and $m_Z$ and $m_W$ are (in lowest order) the physical masses of the $Z$ and
  $W$ bosons, respectively.  For the charged fields this can be easily
  achieved by a rescaling
\bea
\label{eq-wprimmass}
m_W^2 & = & \left(\frac{1 + h_{\varphi}^{(1)}/2}{1 - h_{\varphi
      W}}\right) m_W^{\prime \,2}\\
 & = & \left(\frac{1 + h_{\varphi}^{(1)}/2}{1 - h_{\varphi W}}\right)
 \frac{g^2 v^2}{4},\nonumber\\[.2cm]
\label{eq-wprimfield} 
W^{\pm}_{\mu} & = & \sqrt{1 - h_{\varphi W}}\;W^{\prime \pm}_{\mu}.
\eea
In the approximation linear in the anomalous couplings (\ref{eq-wprimmass})
agrees with~(4.5a) in~\cite{Buchmuller:1985jz} (where the definition of~$v$
differs by a factor of~$\sqrt{2}$ from ours) and with~(3)
in~\cite{Leung:1984ni}.  In the case of the neutral fields we perform a linear
transformation
\be
\label{eq-vtraf}
\boldsymbol{V}'_{\mu}=C\,\boldsymbol{V}_{\mu},
\ee
where
\be
\boldsymbol{V}_{\mu} = (Z_{\mu}, A_{\mu})^{\rm T}.
\ee
Choosing the non-orthogonal matrix
\be
\label{eq-trafomat}
C = \left( \begin{array}{*{2}{c}}
     \sqrt{d/t} & 0\\
     - b / \sqrt{d t} & 1 / \sqrt{d}
   \end{array} \right)
\ee
with $t = ad - b^2$, we obtain the desired form
\be
\label{eq-diag} 
T = C^{\rm T} T' C = \mathbbm{1},\;\;\;\;\;\;\;\;\;\;
M = C^{\rm T} M' C =  
\left(
\begin{array}{*{2}{c}}
     \displaystyle m_Z^2 & 0\\
     0 & 0
    \end{array} \right),
\ee
where $\mathbbm{1}$ denotes the \mbox{2$\times$2} unit matrix and the squared
physical mass of the $Z$ boson is
\bea
\label{eq-zmass}
m_Z^2 & = & \frac{d}{t} \left(1 + \frac{1}{2}
     \left(h_{\varphi}^{(1)} + h_{\varphi}^{(3)}\right)\right)
     m_Z^{\prime \,2}\\
 & = & \frac{d}{t} \left(1 + \frac{1}{2}
     \left(h_{\varphi}^{(1)} + h_{\varphi}^{(3)}\right)\right)
     \frac{g^2 + g^{\prime \, 2}}{4} v^2.\nonumber
\eea
We remark that the simultaneous diagonalisation of the kinetic and mass terms
in the neutral gauge-boson sector is completely analogous to the introduction
of normal coordinates in the problem of small oscillations in mechanics, see
for instance~\cite{bib:gold}.  This kind of
diagonalisation~(\ref{eq-trafomat}) has been done in~\cite{Kuroda:1986bj},
where the mixing term of a $W_3$ and a photon field is studied.  A similar
procedure is performed in~\cite{Burgess:1993vc} where operators up to
dimension five are considered.  In the approximation linear in the anomalous
couplings~(\ref{eq-zmass}) agrees with~(4.5b) in~\cite{Buchmuller:1985jz} and
with~(4) in~\cite{Leung:1984ni}.

Similarly to the gauge bosons we now consider the terms of the Lagrangian
quadratic in the Higgs field
\bea
\mathscr{L}^{(2)}_H & = & \frac{1}{2} \left(1 + \frac{1}{2}
  \left(h_{\varphi}^{(1)} + h_{\varphi}^{(3)}\right)\right)
\left(\partial_{\mu} H'\right) \left(\partial^{\;\!\mu} H'\right)\nonumber\\
 & & \mbox{} - \frac{1}{2} m_H^{\prime \, 2} H^{\prime \, 2},
\eea
where $m_H^{\prime \, 2}$ is given by~(\ref{eq-Higgsmass}).  To obtain the
standard form
\be
\mathscr{L}^{(2)}_H = \frac{1}{2} \left(\partial_{\mu} H\right)
\left(\partial^{\;\!\mu} H\right) - \frac{1}{2} m_H^2 H^2,
\ee
we define the physical Higgs-boson mass and physical Higgs field by a
rescaling
\bea
m_H^2 & = & \frac{m_H^{\prime \, 2}}{1 + \big(h_{\varphi}^{(1)} +
  h_{\varphi}^{(3)}\big)/2}\\[.25cm]
\label{eq-renHiggs}
H & = & \sqrt{1 + \big(h_{\varphi}^{(1)} + h_{\varphi}^{(3)}\big)/2}\;\; H'.
\eea
For the original Higgs-doublet field in the unitary gauge we find
from~(\ref{eq-Higgs}) and~(\ref{eq-renHiggs})
\be
\varphi (x) = \frac{1}{\sqrt{2}} \left(\ba{c}0 \\ v + \Big(1 +
  (h_{\varphi}^{(1)} + h_{\varphi}^{(3)})/2\Big)^{-1/2} H(x)\ea\right).
\ee
For non-zero \mbox{$h_{\varphi}^{(1)} + h_{\varphi}^{(3)}$} this differs from
the SM~result.

To analyse the phenomenology of the effective Lagrangian (\ref{eq-Leff}) we
also have to express the dimension-six operators~(\ref{eq-Ow})
to~(\ref{eq-Ophi1}) in terms of the physical fields $W^{\pm}$, $Z$, $A$
and~$H$.  In particular, we have to substitute the Higgs field according
to~(\ref{eq-Higgs}) and~(\ref{eq-renHiggs}).  Due to~(\ref{eq-wprimfield}),
(\ref{eq-vtraf}) and (\ref{eq-trafomat}) the Lagrangian~(\ref{eq-Leff}), and
particularly the \mbox{$\gamma WW$}, \mbox{$ZWW$}, \mbox{$\gamma \gamma WW$}
and \mbox{$\gamma \gamma H$}~vertices depend then on the anomalous couplings
in a non-linear way.  We list these vertices in Sect.~\ref{sec-tgcs} where we
treat the triple- and quartic-gauge couplings in detail.

The diagonalisation has an important consequence concerning the operators
$O_{\varphi W}$ and $O_{\varphi B}$.  Notice that the $v^2$-terms of these
operators are proportional to the gauge invariant kinetic terms of the SM
Lagrangian, see the~first two terms of~(\ref{eq:smlagr}).  Therefore, after
the substitution of the physical fields, these operators do not give rise to
anomalous three- or four-gauge-boson couplings, see Sect.~\ref{sec-tgcs}.
However, these operators contribute to the \mbox{$\gamma \gamma H$} vertex.

In the next section we shall analyse the consequences of the effective
Lagrangian (\ref{eq-Leff}) and of the diagonalisation~(\ref{eq-lv})ff for the
gauge-boson-fermion couplings.

%%%%%%%%%%%%%%%%%%%%%%%%%%%%%%%%%%%%%%%%%%%%%%%%%%%%%%%%%%%%%%%%%
%%%%%%%%%%%%%%%%%%%%%%%%%%%%%%%%%%%%%%%%%%%%%%%%%%%%%%%%%%%%%%%%%

\section[Gauge-boson-fermion interactions and electroweak
parameters]{Gauge-boson-fermion interactions and\\ electroweak parameters}
\label{sec-para}
\setcounter{equation}{0}

The Lagrangian~(\ref{eq-Leff}) contains the two gauge couplings $g$ and~$g'$.
Apart from that it contains two parameters $\mu$ and~$\lambda$ from the Higgs
potential, nine fermion masses, four parameters of the CKM matrix~$V$, and ten
anomalous couplings~$h_i$.  We can express the original parameters~$\mu$
and~$\lambda$ in terms of~$m_H$ and~$v$ according to
\bea
\mu^2 & = & \frac{m_H^{\prime \, 2}}{2} = \frac{1}{2} \left(1 +
  \frac{1}{2}\left(h_{\varphi}^{(1)} + h_{\varphi}^{(3)}\right)\right)
m_H^2,\\
\label{eq-lambdafund}
\lambda & = & \frac{m_H^{\prime \, 2}}{2 v^2} = \frac{1}{2 v^2} \left(1 +
  \frac{1}{2}\left(h_{\varphi}^{(1)} + h_{\varphi}^{(3)}\right)\right) m_H^2.
\eea
We call $g$, $g'$ and~$v$ the electroweak parameters.  We denote the scheme
that uses as input the parameters from the Lagrangian~(\ref{eq-Leff}), but
$m_H$ and $v$ instead of~$\mu$ and~$\lambda$, by~$P_{\mathscr{L}}$, see
Tab.~\ref{tab:parameters}.  The quantities $s'_{\rm w}$, $c'_{\rm w}$ and
$e'$, which are the sine and cosine of the weak mixing angle and the positron
charge if we set all anomalous couplings to zero, are given in terms of the
electroweak parameters in~(\ref{eq-sinepr}), (\ref{eq-cosinepr})
and~(\ref{eq-epr}), and this leads to the standard relations for the
electroweak observables.  However, with non-zero anomalous couplings, that is
with the full Lagrangian~(\ref{eq-Leff}), the relations of the three
parameters $g$, $g'$ and $v$ to observables depend on the anomalous couplings.

In this section we take a look at the gauge-boson-fermion interactions and
introduce two more sets of electroweak input parameters, see
Tab.~\ref{tab:parameters}.
\begin{table*}
\centering
\begin{tabular}{lccc}
\hline
&&&\\[-.35cm]
parameters & $P_{\mathscr{L}}$ scheme & $P_Z$ scheme & $P_W$ scheme\\
\hline
&&&\\[-.25cm]
electroweak & $g$, $g'$, $v$ & $\alpha(m_Z)$, $G_{\rm F}$, $m_Z$ &
$\alpha(m_Z)$, $G_{\rm F}$, $m_W$\\
Higgs-boson mass & $m_H$ & $m_H$ & $m_H$\\
fermion masses & $m_{\rm u}$,\ldots, $m_{\tau}$ & $m_{\rm u}$,\ldots,
$m_{\tau}$ & $m_{\rm u}$,\ldots, $m_{\tau}$\\
4 CKM parameters & $V$ & $V$ & $V$\\
10 anomalous couplings & $h_W$,\ldots, $h_{\varphi}^{(3)}$ &
$h_W$,\ldots, $h_{\varphi}^{(3)}$ &$h_W$,\ldots, $h_{\varphi}^{(3)}$\\[.05cm]
\hline
\end{tabular}
\caption{\label{tab:parameters}Three parameter sets used in the analysis:
  $P_{\mathscr{L}}$, $P_Z$ and $P_W$~schemes.}
\end{table*}
In these schemes, that we call $P_Z$ and $P_W$, we choose in place of $g$,
$g'$ and $v$ as free parameters the fine structure constant at the $Z$~scale,
\mbox{$\alpha(m_Z)$}, Fermi's constant $G_{\rm F}$, and the mass of the $Z$ or
$W$ boson, respectively.
For our numerics we take
\be
\label{eq-numinput}
1 / \alpha (m_Z) = 128.95(5),\;\;
G_{\rm F} = 1.16639(1) \times 10^{-5}\;{\rm GeV}^{-2}
\ee
from Sect.~16.3 of~\cite{Group:2002mc} and from~\cite{Hagiwara:pw},
respectively.  Moreover, from~\cite{Hagiwara:pw}, we use in the $P_Z$~scheme
\be
\label{eq-numzmass}
m_Z = 91.1876(21)\;{\rm GeV},
\ee
and in the $P_W$~scheme
\be
\label{eq-numwmass}
m_W = 80.423(39)\;{\rm GeV}.
\ee
The small errors on the quantities~(\ref{eq-numinput}) to~(\ref{eq-numwmass})
are negligible for our purposes and will be neglected below.  We use as input
parameter \mbox{$\alpha (m_Z)$} and not the more precisely known~\mbox{$\alpha
(0)$}, since most of the observables which we consider below refer to a high
scale of at least~$m_Z$.  In the following we will denote by~$e$ the positron
charge at~$m_Z$,
\be
e = \sqrt{4 \pi \alpha (m_Z)},
\ee
and refer to~$e$ as the physical positron charge.  This is legitimate in
tree-level calculations.  How we include radiative corrections in our
calculations will be discussed in Sect.~\ref{sec-lep} below.

We use the $P_Z$~scheme for all LEP and SLC observables that we consider in
Sect.~\ref{sec-lep}.  In the scheme $P_Z$, one can calculate the $W$~mass
$m_W^{\rm SM}$ in the SM with a certain theoretical accuracy.  Using the
effective Lagrangian~(\ref{eq-Leff}) instead of the SM Lagrangian gives a
different prediction, $m_W$.  Indeed, as we will see in Sect.~\ref{sec-lep},
two anomalous couplings have an impact on $m_W$ in the $P_Z$~scheme.  However,
for our analysis of \mbox{$e^+e^- \rightarrow WW$} in
Sect.~\ref{ssec-tgcsatnlc} the use of the $P_Z$~scheme with $m_W$ depending on
the anomalous couplings is very inconvenient.
In~\cite{Diehl:2002nj,Diehl:2003qz} $m_W$ is assumed to be a fixed
parameter---as is legitimate and usually done in the form-factor
approach---and not expanded in anomalous couplings.  This is for good reason:
a change of $m_W$ changes the kinematics of \mbox{$e^+e^- \rightarrow WW$} and
the reconstruction of the final state.  Therefore, in
Sect.~\ref{ssec-tgcsatnlc} we use the $P_W$~scheme with $m_W$ instead of $m_Z$
as input.  In this case the $Z$~mass is a parameter that depends on the
anomalous couplings~$h_i$.

Next we consider the fermion-gauge-boson-interaction part $\mathscr{L}_{\rm
int}$ of the Lagrangian (\ref{eq-Leff}).  Since we have not explicitly added
any gauge-boson-fermion operators we get---in the original parameters---the SM
expression.  In terms of the fields $A_{\mu}'$, $Z_{\mu}'$ and
$W_{\mu}^{\prime \pm}$, (\ref{eq-prfielddef1}) and~(\ref{eq-prfielddef2}), we
have thus (see (22.77), (22.123) of~\cite{Nachtmann:ta})
\bea
\label{eq-lint1}
\mathscr{L}_{\rm int} & = & - e' \bigg( A_{\mu}' \mathcal{J}^{\mu}_{\rm em} + 
\frac{1}{s_{\rm w}' c_{\rm w}'}Z_{\mu}' \mathcal{J}^{\prime\mu}_{\rm NC}\\
 & & \mbox{} + \frac{1}
{\sqrt{2}s_{\rm w}'} \left(W^{\prime +}_{\mu} \mathcal{J}^{\mu}_{\rm CC} 
+ {\rm H.c.} \right) \bigg)\nonumber
\eea
with the SM currents
\bea
\label{eq-defemc}
\mathcal{J}^{\mu}_{\rm em}&=& \overline{\psi} \gamma^{\mu}
(\mathbf{T}_3 + \mathbf{Y}) \psi, \\
\label{eq-defnc}
\mathcal{J}^{\prime\mu}_{\rm NC} & = &  \overline{\psi} \gamma^{\mu}
\mathbf{T}_3 \psi - s^{\prime \, 2}_{\rm w} \mathcal{J}^{\mu}_{\rm em},\\
\mathcal{J}^{\mu}_{\rm CC} & = & \overline{\psi} \gamma^{\mu}
(\mathbf{T}_1 + i \mathbf{T}_2) \psi.
\eea
Here $\psi$ is the spinor for all lepton and quark fields.  With the mere SM
Lagrangian, $e'$ is the physical positron charge.  Including the dimension-six
operators we can express the interaction terms through the physical fields
using~(\ref{eq-wprimfield}) to~(\ref{eq-trafomat}):
\bea
\label{eq-Lint2}
\mathscr{L}_{\rm int} & = & - e \bigg( A_{\mu}
\mathcal{J}^{\mu}_{\rm em} + G_{\rm NC} Z_{\mu}\mathcal{J}^{\mu}_{\rm NC}\\
 & & \mbox{} + G_{\rm CC} \left(W^+_{\mu} \mathcal{J}^{\mu}_{\rm CC} + {\rm
     H.c.} \right)\bigg),\nonumber
\eea
where the physical positron charge (at the $Z$~scale) is given by
\be
\label{eq-poscharge}
e = \sqrt{4 \pi \alpha (m_Z)} = \frac{e'}{\sqrt{d}},
\ee
and the physical neutral current by
\be
\label{eq-neutrc}
\mathcal{J}^{\mu}_{\rm NC} = \overline{\psi} \gamma^{\mu} \mathbf{T}_3 \psi -
s^2_{\rm eff} \mathcal{J}^{\mu}_{\rm em} 
\ee
with
\be
\label{eq-sineffan}
s_{\rm eff}^2 \equiv \sin^2 \theta^{\rm lept}_{\rm eff} = s^{\prime \,
2}_{\rm w} + \frac{b}{d} s'_{\rm w} c'_{\rm w}. 
\ee
The neutral- and charged-current couplings are
\be
\label{eq-nccc}
G_{\rm NC} = \frac{1}{s'_{\rm w} c'_{\rm w}}
\frac{d}{\sqrt{t}},\;\;\;\;\;\;\;\;G_{\rm CC} = \frac{1}{\sqrt{2}
s'_{\rm w}} \frac{\sqrt{d}}{\sqrt{1 - h_{\varphi W}}}.
\ee

The electromagnetic, the neutral- and the charged-current interactions are
modified by the anomalous couplings in a universal way for fermions with the
same quantum numbers.  With our definition~(\ref{eq-sineffan}) of the
effective leptonic weak mixing angle the neutral current~(\ref{eq-neutrc}) has
the same form as in the SM, cf.~(\ref{eq-defnc}).  We write the neutral
current as
\be
\label{eq-gagv}
\mathcal{J}^{\mu}_{\rm NC}= \sum_{\rm f} \frac{1}{2}
\overline{\rm f}\left(g_{\rm V}^{\rm f} \gamma^{\mu} - g_{\rm A}^{\rm f}
  \gamma^{\mu} \gamma_5\right) {\rm f},  
\ee
where f denotes any fermion.  Then we find for the vector and axial-vector
neutral-current couplings of leptons
\be
\label{eq-vecax}
g_V^{\ell} = 2 s_{\rm eff}^2 - \frac{1}{2},\;\;\;\;\;\;\;\;
g_A^{\ell} = - \frac{1}{2},
\ee
with $\ell = e, \mu,\tau$.  Using~(\ref{eq-vecax}), we find the usual
expression for~$s_{\rm eff}^2$ \cite{Haywood:1999qg}
\be
\label{eq-sineff}
\sin^2 \theta^{\rm lept}_{\rm eff} = \frac{1}{4} \bigg(
1-\frac{g_{\rm V}^{\ell}} {g_{\rm A}^{\ell}}\bigg).
\ee

Fermi's constant is given by two charged-current interactions in the low
energy limit where the $W$-boson propagator becomes point-like, see e.g.\
Sect.~22.3 of~\cite{Nachtmann:ta}:
\be
\label{eq-gfdef}
G_{\rm F} = \frac{\sqrt{2} e^2}{4 m_W^2} G_{\rm CC}^2.
\ee
It is related to the vacuum expectation value~$v$ of the original Higgs
field~$\varphi$, see (\ref{eq-Higgs}), through
\be
\label{eq-vexp}
v = \left(\sqrt{2} G_{\rm F}\right)^{-1/2} \left(1 +
  h_{\varphi}^{(1)}/2\right)^{-1/2}.
\ee
This is obtained by inserting in~(\ref{eq-gfdef}) for $e$, $G_{\rm CC}$ and
$m_W$ the expressions following from~(\ref{eq-poscharge}), (\ref{eq-nccc})
and~(\ref{eq-wprimmass}), respectively.  For \mbox{$h_{\varphi}^{(1)} = 0$},
(\ref{eq-vexp}) becomes the tree-level SM relation between $v$ and~$G_{\rm
F}$.  The parameter~$\lambda$ from the Higgs potential is therefore,
cf.~(\ref{eq-lambdafund}),
\be
\lambda = \frac{G_{\rm F} m_H^2}{\sqrt{2}} \left(1 + \frac{1}{2}
  \left(h_{\varphi}^{(1)} + h_{\varphi}^{(3)}\right)\right) \left(1 +
    h_{\varphi}^{(1)}/2\right).
\ee
In the following two subsections we determine how the remaining original
parameters of the Lagrangian~(\ref{eq-Leff}) are related to our input
parameters in the $P_Z$ and $P_W$ schemes.  Knowing these relations one can
express all constants in the Lagrangian by either of the two electroweak
parameter sets plus the anomalous couplings~$h_i$.

%%%%%%%%%%%%%%%%%%%%%%%%%%%%%%%%%%%%%%%%%%%%%%%%%%%%%%%%%%%%%%%%%

\subsection[$P_Z$ scheme]{\boldm{P_Z} scheme}
\label{ssec-pz}

We now show how the original parameters in the effective
Lagrangian~(\ref{eq-Leff}), are expressed by the input parameters of the
$P_Z$~scheme, see Tab.~\ref{tab:parameters}.  The physical $Z$ mass $m_Z$ and
\mbox{$\alpha (m_Z)$} are given in terms of the $P_{\mathscr{L}}$~parameters
in~(\ref{eq-zmass}) and~(\ref{eq-poscharge}), respectively.  In the
$P_Z$~scheme the $W$~mass $m_W$ is a derived quantity.  The relation of $m_W$
to the $P_{\mathscr{L}}$~parameters is given in~(\ref{eq-wprimmass}).  We
use~(\ref{eq-wprimmass}), the relation \mbox{$m'_W = c'_{\rm w} m'_Z$} and
express $m'_Z$ by means of~(\ref{eq-zmass}) to obtain the tree-level result
for the squared $W$~mass in the framework of the effective
Lagrangian~(\ref{eq-Leff}):
\be
\label{eq-mwexact}
m_W^2 = \frac{t}{d} \frac{1 + h_{\varphi}^{(1)}/2}{\Big(1 -
h_{\varphi W}\Big)\left(1 + \big(h_{\varphi}^{(1)} +
  h_{\varphi}^{(3)}\big)/2\right)} c^{\prime \, 2}_{\rm w} m^2_Z,
\ee
Inserting (\ref{eq-nccc}) and (\ref{eq-mwexact}) in~(\ref{eq-gfdef}) we obtain
an equation for~$s'_{\rm w}$:
\be
\label{eq-swexact}
s_{\rm w}^{\prime \, 2} = \frac{1}{2} \left\{1 - \sqrt{1 - \frac{e^2}{\sqrt{2}
        G_{\rm F} m^2_Z} \frac{d^2}{t} \frac{1 + \big(h_{\varphi}^{(1)} +
        h_{\varphi}^{(3)}\big)/2}{1 + h_{\varphi}^{(1)}/2}} \right\}.
\ee
Note that $d$ and $t$ contain $s'_{\rm w}$ and $c'_{\rm w}$,
see~(\ref{eq-adef}) to~(\ref{eq-ddef}).  Therefore (\ref{eq-swexact}) is only
an implicit equation for $s'_{\rm w}$, which is not easy to solve exactly.  We
denote the right hand side of~(\ref{eq-swexact}) for the case where all
anomalous couplings are set to zero by~$s_0^2$:
\be
\label{eq-s0def}
s_0^2 \equiv \frac{1}{2} \left(1 - \sqrt{1 - \frac{e^2}{\sqrt{2} G_{\rm F}
      m_Z^2}} \right),\;\;\;\;\;\;\;\;\;\;c_0^2 \equiv 1 - s_0^2.
\ee
Hence $s_0$ and $c_0$ are not independent parameters but combinations of input
parameters in the $P_Z$~scheme.  In the SM, they are identical to the sine and
cosine of the weak mixing angle.  To linear order in the anomalous couplings
we obtain from~(\ref{eq-swexact}) in the $P_Z$~scheme
\bea
\label{eq-swdef}
s_{\rm w}^{\prime \, 2} & = & s_0^2 \bigg(1 + c_0^2 \left(h_{\varphi W} -
  h_{\varphi B}\right) + \frac{4 s_0 c_0^3}{c_0^2 - s_0^2} h_{W \mskip -3mu
  B}\nonumber\\
 & & \mbox{} + \frac{c_0^2}{2 \left(c_0^2 - s_0^2\right)}
  h_{\varphi}^{(3)}\bigg).
\eea
Expanding~(\ref{eq-sineffan}) to first order in the couplings we find in the
$P_Z$~scheme
\be
\label{eq-seffdef}
s_{\rm eff}^2 = s_0^2 \left(1 + \frac{c_0}{s_0 (c_0^2 - s_0^2)}
h_{W \mskip -3mu B} + \frac{c_0^2}{2 (c_0^2 - s_0^2)}
h_{\varphi}^{(3)}\right).
\ee
Using~(\ref{eq-swdef}) and~(\ref{eq-seffdef}) the quantities $s'_{\rm w}$,
$c'_{\rm w}$ and $s_{\rm eff}^2$ in~(\ref{eq-neutrc}) and~(\ref{eq-nccc}) can
be expressed as functions of $s_0$ and anomalous couplings in the linear
approximation.  The neutral- and charged-current couplings~(\ref{eq-nccc})
read to first order in the anomalous couplings in the $P_Z$ scheme
\bea
\label{eq-nccoupl}
\!\!\!\!\!\!\!\!\!\!\!\!\! G_{\rm NC} & = & \frac{1}{s_0 c_0} \left(1 -
\frac{1}{4} h_{\varphi}^{(3)}\right),\\
\label{eq-cccoupl}
\!\!\!\!\!\!\!\!\!\!\!\!\! G_{\rm CC} & = & \frac{1}{\sqrt{2} s_0} \bigg(1 + \frac{s_0
  c_0}{s_0^2 - c_0^2} h_{W \mskip -3mu B} + \frac{c_0^2}{4(s_0^2 -
c_0^2)} h_{\varphi}^{(3)}\bigg).
\eea
For non-zero anomalous couplings an exact result for the $W$-boson mass is, in
principle, obtained by inserting the solution for $s'_{\rm w}$
from~(\ref{eq-swexact}) into~(\ref{eq-mwexact}).  Expanding to first order in
the anomalous couplings we obtain in the $P_Z$ scheme
\be
\label{eq-wmass}
m_W = c_0 m_Z \left(1 + \frac{s_0 c_0}{s_0^2 - c_0^2} h_{W \mskip -3mu B} +
  \frac{c_0^2}{4 \left(s_0^2 - c_0^2\right)} h_{\varphi}^{(3)}\right).
\ee
This equation is a relation at tree-level.  The way in which radiative
corrections are taken into account in our analysis is explained at the
beginning of Sect.~\ref{sec-lep}.  For the vacuum expectation value~$v$ we
obtain to linear order in the anomalous couplings in the $P_Z$~scheme,
expanding in~(\ref{eq-vexp})
\be
\label{eq-vexpz}
v = \left(\sqrt{2} G_{\rm F}\right)^{-1/2} \left(1 -
  h_{\varphi}^{(1)}/4\right).
\ee

%%%%%%%%%%%%%%%%%%%%%%%%%%%%%%%%%%%%%%%%%%%%%%%%%%%%%%%%%%%%%%%%%

\subsection[$P_W$ scheme]{\boldm{P_W} scheme}
\label{ssec-pw}

Similarly as in the preceding subsection we now express various quantities in
the $P_W$~scheme, see Tab.~\ref{tab:parameters}.  Inserting~(\ref{eq-nccc})
into~(\ref{eq-gfdef}) and solving for $s_{\rm w}^{\prime \, 2}$ we obtain
\be
\label{eq-sww}
s_{\rm w}^{\prime \, 2} = \frac{e^2}{4 \sqrt{2} G_{\rm F} m_W^2}\frac{d}{1 -
  h_{\varphi W}}.
\ee
Notice that in this equation $d$ contains~$s'_{\rm w}$ and~$c'_{\rm w}$.
Therefore it is only an implicit equation for $s_{\rm w}^{\prime \, 2}$
like~(\ref{eq-swexact}).  For the case where all $h_i$ are zero the right hand
side of~(\ref{eq-sww}) is given by
\be
\label{eq-s1def}
s^2_1 \equiv \frac{e^2}{4 \sqrt{2} G_{\rm F} m_W^2},\;\;\;\;\;\;\;\;\;\;c_1^2
\equiv 1 - s_1^2. 
\ee
Here $s_1$ and $c_1$ are combinations of input parameters of~$P_W$.
Expanding~(\ref{eq-sww}) to linear order in the anomalous couplings we obtain
in the $P_W$~scheme
\be
s_{\rm w}^{\prime \, 2} = s^2_1 \left(1 + c_1^2 \left(h_{\varphi W} -
    h_{\varphi B}\right) + 2 s_1 c_1 h_{W \mskip -3mu B}\right).
\ee
We expand~(\ref{eq-sineffan}) to first order in the~$h_i$:
\be
\label{eq-seffpwscheme}
s_{\rm eff}^2 = s_1^2 \left(1 + \frac{c_1}{s_1} h_{W \mskip -3mu B}\right).
\ee
For the neutral-current coupling~(\ref{eq-nccc}) we find to first order in the
anomalous couplings in~$P_W$
\be
G_{\rm NC} = \frac{1}{s_1 c_1} \left(1 + \frac{s_1}{c_1} h_{W
\mskip -3mu B}\right).
\ee
Here due to~(\ref{eq-gfdef}) and~(\ref{eq-s1def}) the charged-current coupling
is given exactly by
\be
\label{eq-ccpw}
G_{\rm CC} = \frac{1}{\sqrt{2} s_1},
\ee
and not modified by anomalous couplings.  Using the relation \mbox{$m'_Z =
m'_W / c'_{\rm w}$} as well as~(\ref{eq-wprimmass}) and~(\ref{eq-zmass}) we
find for the squared $Z$~mass in~$P_W$
\be
m_Z^2 = \frac{d}{t} \frac{\left(1 + \big(h_{\varphi}^{(1)} +
h_{\varphi}^{(3)}\big)/2\right) \Big(1 - h_{\varphi W}\Big)}{1 +
h_{\varphi}^{(1)}/2} \frac{m_W^2}{c_{\rm w}^{\prime \, 2}},
\ee
where for $s'_{\rm w}$ in $d$ and $t$ the solution to~(\ref{eq-sww}) has to be
inserted, and~\mbox{$c'_{\rm w} = \sqrt{1 - s_{\rm w}^{\prime \, 2}}$}.  So
far this is an exact expression for~$m_Z$.  To first order in the $h_i$ the
$Z$~mass is
\be
\label{eq-pwzmass}
m_Z = \frac{m_W}{c_1} \left(1 + \frac{s_1}{c_1} h_{W \mskip -3mu B} +
  \frac{1}{4} h_{\varphi}^{(3)}\right).
\ee
For the vacuum expectation value $v$ to linear order in the $h_i$ we have the
same expression as in the $P_Z$~scheme,~(\ref{eq-vexpz}).

%%%%%%%%%%%%%%%%%%%%%%%%%%%%%%%%%%%%%%%%%%%%%%%%%%%%%%%%%%%%%%%%%
%%%%%%%%%%%%%%%%%%%%%%%%%%%%%%%%%%%%%%%%%%%%%%%%%%%%%%%%%%%%%%%%%

\section{Limits from LEP and SLC}
\label{sec-lep}
\setcounter{equation}{0}

In this section we discuss the impact of the additional operators on precision
observables measured at LEP and SLC.  As mentioned before we use the
$P_Z$~scheme in the entire Sect.~\ref{sec-lep}.  Our procedure is as follows:
We calculate the tree-level prediction~$X_{\rm tree}$ of an observable in the
framework of the effective Lagrangian~(\ref{eq-Leff}).  Then~$X_{\rm tree}$
can be expanded to first order in~$h_i$
\be
\label{eq-xtree}
X_{\rm tree} = X_{\rm tree}^{\rm SM} \left(1 + \sum_i h_i \hat{X}_i\right),
\ee
where $X_{\rm tree}^{\rm SM}$ is the result if we set all anomalous couplings
to zero, that is the result one obtains from the tree-level calculation with
the mere SM Lagrangian.  At higher loop-order both~$X_{\rm tree}$ and~$X_{\rm
tree}^{\rm SM}$ receive corrections.  It is well known how to calculate
radiative corrections in the~SM, see for instance~\cite{Bohm:yx}.  As already
mentioned in the introduction radiative corrections can also be evaluated for
a non-renormalisable Lagrangian like ours in~(\ref{eq-Leff}) using the
effective-field-theory techniques, see for instance~\cite{Weinberg:mt}.  This
would result in a renormalisation of the original anomalous couplings and in
the introduction of further anomalous terms of higher dimension with free
coefficients.  Thus, radiative corrections to our anomalous couplings should
only give terms having further suppression factors $\alpha$ and/or \mbox{$(v /
\Lambda)$} and will be neglected in the following.  In detail, we expand the
complete result $X$ for an observable as
\be
\label{eq-xexact}
X = X^{\rm SM} \left(1 + \sum_i h_i \hat{X}_i\right) + \Delta \widetilde{X},
\ee
where $X^{\rm SM}$ is the complete SM result and the $\hat{X}_i$ are the {\em
  same} expressions as in~(\ref{eq-xtree}).  The term \mbox{$\Delta
  \widetilde{X}$} contains then radiative corrections times and to anomalous
  couplings and will be neglected in the following.  To get bounds on
  the~$h_i$ we insert the experimental values for~$X$ and use the well-known
  higher-order results for~$X^{\rm SM}$.  The linear parts~$\hat{X}_i$ are
  obtained from the tree-level expansion~(\ref{eq-xtree}).  The experimental
  errors~\mbox{$\delta X$} together with the theoretical
  uncertainties~\mbox{$\delta X^{\rm SM}$} of the SM calculation allow us then
  to derive bounds on the~$h_i$.  The theoretical values~$X^{\rm SM}$ depend
  on the unknown Higgs mass~$m_H$, see~\cite{Group:2002mc}, and we shall
  discuss the bounds as functions of~$m_H$.

As first observable we consider the leptonic mixing angle~(\ref{eq-sineffan})
for which we get in the $P_Z$ scheme~(\ref{eq-seffdef}).  There we can
identify $s_0$ from~(\ref{eq-s0def}) as the tree-level SM result
\be
\left. s^{\rm SM}_{\rm eff}\right|_{\rm tree} = s_0.
\ee
According to~(\ref{eq-xexact}) and~(\ref{eq-seffdef}) we set now
\bea
s_{\rm eff}^2 & = & \left(s^{\rm SM}_{\rm eff}\right)^2 \left(1 +
  \frac{c_0}{s_0 (c_0^2 - s_0^2)} h_{W \mskip -3mu B} + \frac{c_0^2}{2 (c_0^2
  - s_0^2)} h_{\varphi}^{(3)}\right)\nonumber\\[.2cm]
\label{eq-seffexpansion}
 & = & \left(s_{\rm eff}^{\rm SM}\right)^2 \left(1 + 3.39 h_{W \mskip -3mu B}
  + 0.71 h_{\varphi}^{(3)}\right).
\eea
Here $s^{\rm SM}_{\rm eff}$ is the leptonic mixing angle in the SM, including
radiative corrections, and the numerical values are obtained
with~(\ref{eq-numinput}) and~(\ref{eq-numzmass}).

The partial widths of the $Z$ into a pair of fermions calculated from the
Lagrangian~(\ref{eq-Leff}) on tree-level are
\be
\label{eq-partial}
\left. \Gamma_{\rm f \overline{f}}\right|_{\rm tree} = \frac{e^2 m_Z}{48 \pi}
G_{\rm NC}^2 N_c^{\rm f} \chi_{\rm f},
\;\;\;\;\;
\chi_{\rm f} = \left( g_{\rm V}^{\rm f} \right)^2 + \left( g_{\rm A}^{\rm f}
\right)^2,
\ee
where $N_c^{\rm f} = 1$ for leptons and  $N_c^{\rm f} = 3$ for quarks.  For
neutrinos, charged leptons, up- and down-type quarks we get, respectively,
\begin{alignat}{4}
\label{eq-chidef1}
&\chi_{\nu} && = \frac{1}{2},&\;\;\;
&\chi_{\ell} && = \frac{1}{2} - 2 s_{\rm eff}^2 + 4 s_{\rm eff}^4,\\
\label{eq-chidef2}
&\chi_{\rm u} && = \frac{1}{2} - \frac{4}{3} s_{\rm eff}^2 +
\frac{16}{9} s_{\rm eff}^4,&\;\;\;
&\chi_{\rm d} && = \frac{1}{2} - \frac{2}{3} s_{\rm eff}^2 +
\frac{4}{9} s_{\rm eff}^4.
\end{alignat}
In~(\ref{eq-partial}) we have neglected all fermion masses.  Setting all
anomalous couplings to zero we find expressions for the tree-level partial
widths in the SM as in Chapter~25 of~\cite{Nachtmann:ta}.  The partial widths
in (\ref{eq-partial}) depend on the anomalous couplings through $G_{\rm
NC}$~(\ref{eq-nccoupl}) and through~$s_{\rm eff}^2$ in~$\chi_{\rm f}$.
Expanding (\ref{eq-partial}) to first order in the anomalous couplings and
using our prescription~(\ref{eq-xexact}), we obtain the following results for
the invisible partial width, the width into one pair of charged leptons
\mbox{$e^+ e^-$}, \mbox{$\mu^+ \mu^-$} or \mbox{$\tau^+ \tau^-$}, the hadronic
and the total widths:
\bea
\Gamma_{\rm inv} & = & \Gamma_{\rm inv}^{\rm SM} \left( 1 -
  \frac{h_{\varphi}^{(3)}}{2}\right),\\[.3cm]
\Gamma_{\ell \ell} & = & \Gamma_{\ell \ell}^{\rm SM} \bigg(1 +
\frac{4 s_0 c_0 (4 s_0^2 - 1) h_{W \mskip -3mu B}}{1 - 6 s_0^2 + 16 s_0^4 -
  16 s_0^6} \\
 & & \mbox{} + \frac{\left(-1 + 2 s_0^2 + 4 s_0^4 \right)
   h_{\varphi}^{(3)}}{2 - 4 s_0^2 (3 - 8 s_0^2 + 8
   s_0^4)}\bigg),\nonumber\\[.3cm]
\Gamma_{\rm had} & = & \Gamma_{\rm had}^{\rm SM} \bigg(1 + \frac{4
s_0 c_0 (44 s_0^2 - 21) h_{W \mskip -3mu B}}{45 - 174 s_0^2 + 256 s_0^4 - 176
s_0^6}\nonumber\\
 & & \mbox{} + \frac{\left(-45 + 90 s_0^2 + 4 s_0^4\right)
   h_{\varphi}^{(3)}}{90 - 348 s_0^2 + 512 s_0^4 - 352
   s_0^6}\bigg),\\[.3cm]
\Gamma_Z & = & \Gamma_Z^{\rm SM} \bigg(1 + \frac{40 s_0 c_0 (8
s_0^2 - 3) h_{W \mskip -3mu B}}{63 - 246 s_0^2 + 400 s_0^4 -320
s_0^6}\nonumber\\
 & & \mbox{} + \frac{\left(-63 + 126 s_0^2 + 40 s_0^4\right)
   h_{\varphi}^{(3)}}{126 - 492 s_0^2 + 800 s_0^4 -640 s_0^6}\bigg).
\eea
Using~(\ref{eq-numinput}), (\ref{eq-numzmass}) and~(\ref{eq-s0def}) we get
numerically
\begin{alignat}{2}
\label{eq-gammainvnum}
&\Gamma_{\rm inv} && = \;\Gamma_{\rm inv}^{\rm SM} (1 - 0.50
h_{\varphi}^{(3)}),\\[.3cm]
&\Gamma_{\ell \ell} && = \;\Gamma_{\ell \ell}^{\rm SM} (1 - 0.47 h_{W
\mskip -3mu B} - 0.60 h_{\varphi}^{(3)}),\\[.3cm]
&\Gamma_{\rm had} && = \;\Gamma_{\rm had}^{\rm SM} (1 - 1.12 h_{W
\mskip -3mu B} - 0.74 h_{\varphi}^{(3)}),\\[.3cm]
\label{eq-gammaznum}
&\Gamma_Z && = \;\Gamma_Z^{\rm SM} (1 - 0.82 h_{W \mskip -3mu B}
 - 0.67 h_{\varphi}^{(3)}).
\end{alignat}
Notice that $s_{\rm eff}^2$, $\Gamma_{\ell \ell}$, $\Gamma_{\rm had}$ and
$\Gamma_Z$ all depend on the couplings $h_{W \mskip -3mu B}$ and
$h_{\varphi}^{(3)}$ in a different way.  In contrast, at tree-level in the~SM
as well as with the Lagrangian~(\ref{eq-Leff}) the hadronic pole cross section
$\sigma^0_{\rm had}$ as well as $R^0_{\ell}$, $R^0_{\rm b}$ and $R^0_{\rm c}$
\cite{Group:2002mc} depend only on $s_{\rm eff}^2$ since they are defined in
terms of ratios of the partial and total widths, such that the anomalous
couplings enter only through the quantities $\chi_{\rm f}$,
see~(\ref{eq-partial}) to~(\ref{eq-chidef2}):
\be
\sigma^0_{\rm had} = \frac{12\pi}{m_Z^2}\frac{\Gamma_{\rm ee}
\Gamma_{\rm had}}{\Gamma_Z^2},
\ee
\vspace{.2cm}
\be
R^0_{\ell} = \Gamma_{\rm had}/\Gamma_{\ell \ell},\;\;\;\;\;\;
R^0_{\rm b} = \Gamma_{\rm b \overline{b}}/\Gamma_{\rm
had},\;\;\;\;\;\;
R^0_{\rm c} = \Gamma_{\rm c \overline{c}}/\Gamma_{\rm
had}.
\ee
Note the deviating definition of the leptonic ratio where $\Gamma_{\rm had}$
appears in the numerator.  Also another group of observables, the quantities
\be
\mathcal{A}_{\rm f} = 2 g_{\rm V}^{\rm f} g_{\rm A}^{\rm f} /
\chi_{\rm f},
\ee
and the forward-backward asymmetries
\be
A_{\rm FB}^{0, \rm f} = \frac{3}{4} \mathcal{A}_{\rm e}
\mathcal{A}_{\rm f},
\ee
are solely functions of $s_{\rm eff}^2$:
\begin{alignat}{4}
&\mathcal{A}_{\nu} && = 1,&\;\;\;
&\mathcal{A}_{\ell} && = \left( \frac{1}{2} - 2 s_{\rm eff}^2 \right)
/ \chi_{\ell},\\
\label{eq-Aquarks}
&\mathcal{A}_{\rm u} && = \left( \frac{1}{2} - \frac{4}{3} s_{\rm
eff}^2 \right) / \chi_{\rm u},&\;\;\;
&\mathcal{A}_{\rm d} && = \left( \frac{1}{2} - \frac{2}{3} s_{\rm
eff}^2 \right) / \chi_{\rm d}.
\end{alignat}
We thus find that a large number of the observables listed in the summary
table~16.1 of~\cite{Group:2002mc} with the combined results from LEP1, SLC,
LEP2 and $W$-boson measurements depends on the anomalous couplings only through
$s_{\rm eff}^2$, that is only through the linear combination in~(\ref{eq-seffexpansion}).
These are the observables
\begin{alignat}{1}
\label{eq-seffobs1}
&\mathcal{A}_{\ell}(\mathcal{P}_{\tau}),\; \mathcal{A}_{\ell}({\rm SLD}),\;
A_{\rm FB}^{0, \ell},\; s_{\rm eff}^2 (\langle Q_{\rm FB}\rangle),\; A_{\rm
  FB}^{0, \rm b},\; A_{\rm FB}^{0, \rm c},\\[.2cm]
\label{eq-seffobs2}
&\Gamma_{\rm inv}/\Gamma_{\ell \ell},\; R^0_{\rm b},\; R^0_{\rm c},\;
\mathcal{A}_{\rm b},\; \mathcal{A}_{\rm c},\\[.2cm]
\label{eq-seffobs3}
&\sigma^0_{\rm had},\; R^0_{\ell}.
\end{alignat}
Their functional dependence on~$s_{\rm eff}^2$ is at tree-level the same for
the Lagrangian~(\ref{eq-Leff}) as in the~SM.  Thus, neglecting again radiative
corrections times and to anomalous couplings, we can use the determination
of~$s_{\rm eff}^2$ from \cite{Group:2002mc} directly for our purposes.  From
the six observables (\ref{eq-seffobs1}) the following value for~$s_{\rm
eff}^2$ is extracted in Tab.~15.4 of~\cite{Group:2002mc}:
\be
\label{eq-seffres}
s_{\rm eff}^2 = 0.23148 \pm 0.00017.
\ee
The errors of the observables~(\ref{eq-seffobs2}) are much larger than those
of the observables~(\ref{eq-seffobs1}) and therefore do not affect this result
within rounding errors, which we have checked explicitly using the tree-level
expressions of the observables~(\ref{eq-seffobs2}).  Among the
observables~(\ref{eq-seffobs1}) the leptonic ones tend to give smaller values
for $s_{\rm eff}^2$ than the hadronic ones.  This has recently been mentioned
in~\cite{Altarelli:2003rm}.  We note that this discrepancy cannot be cured by
the anomalous couplings that we consider in this paper since any choice
for~$h_{W \mskip -3mu B}$ and~$h_{\varphi}^{(3)}$ leads to one particular
value of~$s_{\rm eff}^2$ and the observables depend on~$s_{\rm eff}^2$ as in
the SM.  For the two observables~(\ref{eq-seffobs3}) results are given in
Tab.~2.3 (``with lepton universality'') of~\cite{Group:2002mc}, where they are
correlated with $m_Z$, $\Gamma_Z$ and~$A_{\rm FB}^{0, \ell}$:
\begin{alignat}{2}
\label{eq-mzprec}
m_Z \; [{\rm GeV}] & = &\; 91.1875 &\pm 0.0021,\\
\Gamma_Z \; [{\rm GeV}] & = &\; 2.4952 &\pm 0.0023,\\
\sigma^0_{\rm had} \; [{\rm nb}]& = &\; 41.540 &\pm 0.037,\\
R^0_{\ell} & = &\; 20.767 &\pm 0.025,\\
A_{\rm FB}^{0, \ell} & = &\; 0.0171 &\pm 0.0010.
\end{alignat}
The correlations given in the same table are, in the order $m_Z$, $\Gamma_Z$,
$\sigma^0_{\rm had}$, $R^0_{\ell}$, $A_{\rm FB}^{0, \ell}$,
\be
\label{eq-corrprec}
\left(\ba{rrrrr}
1 & -0.023 & -0.045 & 0.033 & 0.055\\
 & 1 & -0.297 & 0.004 & 0.003\\
 & & 1 & 0.183 & 0.006\\
 & & & 1 & -0.056\\
 & & & & 1\\
\ea\right).
\ee
In our scheme $P_Z$ the $Z$~mass is an input parameter.  The forward-backward
asymmetry $A_{\rm FB}^{0, \ell}$ is already included in the result for~$s_{\rm
eff}^2$ in~(\ref{eq-seffres}).  We thus exclude $m_Z$ and $A_{\rm FB}^{0,
\ell}$ from~(\ref{eq-mzprec}) to~(\ref{eq-corrprec}) by projecting the error
ellipsoid onto the subspace of $\Gamma_Z$, $\sigma^0_{\rm had}$
and~$R^0_{\ell}$.  Since $\Gamma_Z$ depends on the couplings $h_{W \mskip -3mu
B}$ and $h_{\varphi}^{(3)}$ in a different way than~$s_{\rm eff}^2$ we can in
this way extract values on these two couplings from~(\ref{eq-seffres})
to~(\ref{eq-corrprec}).  The SM predictions for $\sigma^0_{\rm had}$,
$R^0_{\ell}$ and in particular for~$\Gamma_Z$ and~$s_{\rm eff}^2$ depend
on~$m_H$.  Their numerical values are taken from Figs.~15.4 and~16.6
of~\cite{Group:2002mc}.  For the convenience of the reader we list these
numbers in Tab.~\ref{tab:smvalues}.
\begin{table}
\centering
\begin{tabular}{lrrrr}
\hline
\hline
 &&&&\\[-.32cm]
$m_H$ \hspace{-.5cm} & 120 GeV & 200 GeV & 500 GeV & $\delta X$\\
\hline
 &&&&\\[-.32cm]
$s_{\rm eff}^2$ \hspace{-.5cm} & 0.23156 & 0.23180 & 0.23230 & 0.00030\\
$\Gamma_Z$ [GeV] \hspace{-.5cm} & 2.4952 & 2.4938 & 2.4902 & 0.0026\\
$\sigma^0_{\rm had}$ [nb] \hspace{-0.5cm} & 41.484 & 41.485& 41.489 & 0.015\\
$R^0_{\ell}$ \hspace{-0.5cm} & 20.737 & 20.732 & 20.723 & 0.018\\
$m_W$ [GeV] \hspace{-0.5cm} & 80.374 & 80.341 & 80.269 & 0.041\\
$\Gamma_W$ [GeV] \hspace{-0.5cm} & 2.0896 & 2.0880 & 2.0832 & 0.0032\\[.02cm]
\hline \hline
\end{tabular}
\caption{\label{tab:smvalues}Values of various observables $X$ predicted by
  the~SM for different Higgs masses.  The dependence of their uncertainties
  \mbox{$\delta X$} on~$m_H$ is negligibly small.  Taken from Figs.~15.4, 16.6
  and~16.9 of~\protect{\cite{Group:2002mc}}.}
\end{table}
In Tab.~\ref{tab:coup1} we list the results for the anomalous couplings
extracted from~(\ref{eq-seffres}), $\Gamma_Z$, $\sigma^0_{\rm had}$
and~$R^0_{\ell}$ for a Higgs mass of 120~GeV, 200~GeV and 500~GeV,
respectively.  The errors include the uncertainties in the SM predictions,
which are mainly due to the uncertainties in \mbox{$\Delta \alpha^{(5)}_{\rm
had} (m_Z^2)$}, \mbox{$\alpha_{\rm s}(m_Z^2)$} and $m_{\rm t}$.
\begin{table}
\centering
\begin{tabular}{lrrrrr}
\hline
\hline
 &&&&&\\[-.31cm]
 && \multicolumn{4}{l}{$s_{\rm eff}^2$,\, $\Gamma_Z$,\, $\sigma^0_{\rm had}$,\,
 $R^0_{\ell}$}\\[.03cm]
\hline
\hline
 &&&&&\\[-.34cm]
$m_H$ \hspace{-.5cm} && 120 GeV & 200 GeV & 500 GeV & $\delta h \times 10^3$\\
\hline
 &&&&&\\[-.32cm]
$h_{W \mskip -3mu B}$ \hspace{-.5cm} & $\times 10^3$ & $-0.26$ & $-0.44$ &
$-0.68$ & 0.81\\
$h_{\varphi}^{(3)}$ \hspace{-.5cm} & $\times 10^3$ & 0.38 & $-0.24$ & $-2.08$
& 2.81\\[.02cm]
\hline \hline
\end{tabular}
\caption{\label{tab:coup1}Prediction of $CP$~conserving couplings in units of
  $10^{-3}$ from the observables listed in the first row.  For~$s_{\rm eff}^2$
  the result~(\protect{\ref{eq-seffres}}) from the
  observables~(\protect{\ref{eq-seffobs1}}) is used.  The results are computed
  for a Higgs mass of 120~GeV, 200~GeV and 500~GeV, respectively.  The errors
  \mbox{$\delta h$} on the couplings and the correlation between the two
  errors are independent of the Higgs mass within rounding errors.  The
  correlation is~$-86\%$.}
\end{table}

We now want to include in the analysis of the anomalous couplings the data of
$W$-mass and -width measurements.  The expansion of~$m_W$ has already been
given in (\ref{eq-wmass}).  For the total width of the $W$~boson we get from
(\ref{eq-Lint2}), (\ref{eq-cccoupl}) and~(\ref{eq-wmass}) at tree-level,
neglecting fermion masses,
\bea
\left. \Gamma_W\right|_{\rm tree} & = & \frac{3 e^2 m_W}{8 \pi} G^2_{\rm CC}
\label{eq-totalw} \\ \nonumber
 & = & \left. \Gamma^{\rm SM}_W\right|_{\rm tree} \bigg( 1 + \frac{3 s_0
     c_0}{s_0^2 - c_0^2} h_{W \mskip -3mu B}\\ 
 & & \mbox{} + \frac{3 c_0^2}{4\left(s_0^2 -
     c_0^2\right)} h_{\varphi}^{(3)}\bigg),\nonumber
\eea
where \mbox{$\left. \Gamma_W^{\rm SM}\right|_{\rm tree} = 3 e^2 c_0 m_Z /(16
\pi s_0^2)$}.  In the $P_Z$~scheme the total width $\Gamma_W$ depends on the
same linear combination of anomalous couplings as $m_W$, see~(\ref{eq-wmass}),
and is three times more sensitive to changes of~$h_{W \mskip -3mu B}$
and~$h_{\varphi}^{(3)}$.  Now we use again our general
prescription~(\ref{eq-xexact}) and insert numerical values for $s_0$ and $c_0$
following from~(\ref{eq-numinput}) and~(\ref{eq-numzmass}).  We obtain then
\begin{alignat}{2}
\label{eq-mwnum}
& m_W && = \;m_W^{\rm SM} (1 - 0.78 h_{W \mskip -3mu B} - 0.36
h_{\varphi}^{(3)}),\\[.3cm]
\label{eq-gammawnum}
& \Gamma_W && = \;\Gamma_W^{\rm SM} (1 - 2.35 h_{W \mskip -3mu B} - 1.07
h_{\varphi}^{(3)}).
\end{alignat}
We recall that in the presence of anomalous couplings all charged-current
interactions are modified in a universal way.  Consequently, we obtain the
same relation~(\ref{eq-totalw}) for all partial widths of the $W$~boson.  The
branching ratios of the $W$~boson are therefore not changed by anomalous
effects, in contrast to those of the $Z$~boson.  We use the experimental
values given in~(16.1) and (16.2) of~\cite{Group:2002mc} derived from LEP,
SPSC and Tevatron data
\begin{alignat}{2}
m_W & = &\;80.449 &\pm 0.034,\\
\Gamma_W & = &\;2.136 &\pm 0.069,
\end{alignat}
where the error correlation is~\mbox{$-6.7\%$}.  Using the SM values for $m_W$
and $\Gamma_W$ from Fig.~16.9 of~\cite{Group:2002mc}, which are shown in
Tab.~\ref{tab:smvalues} for three different Higgs masses, and combining the
bounds from~$m_W$ and~$\Gamma_W$ with the results from Tab.~\ref{tab:coup1} we
get the bounds on the couplings $h_{W \mskip -3mu B}$ and~$h_{\varphi}^{(3)}$
as listed in Tab.~\ref{tab:coup2}.
\begin{table}
\centering
\begin{tabular}{lrrrrr}
\hline
\hline
 &&&&&\\[-.31cm]
 && \multicolumn{4}{l}{$s_{\rm eff}^2$,\, $\Gamma_Z$,\, $\sigma^0_{\rm had}$,\,
 $R^0_{\ell}$,\, $m_W$,\, $\Gamma_W$}\\
\hline
\hline
 &&&&&\\[-.34cm]
$m_H$ \hspace{-.5cm} && 120 GeV & 200 GeV & 500 GeV & $\delta h \times 10^3$\\
\hline
 &&&&&\\[-.32cm]
$h_{W \mskip -3mu B}$ \hspace{-.5cm} & $\times 10^3$ & $-0.04$ & $-0.20$ &
$-0.43$ & 0.79\\
$h_{\varphi}^{(3)}$ \hspace{-.5cm} & $\times 10^3$ & $-1.17$ & $-1.88$ &
$-3.81$ & 2.39\\[.02cm]
\hline
\hline
\end{tabular}
\caption{\label{tab:coup2}Same as Tab.~\protect{\ref{tab:coup1}}, but here
  $m_W$ and $\Gamma_W$ are included as observables.  The correlation of the
  errors is~$-88\%$.}
\end{table}
%
%

%%%%%%%%%%%%%%%%%%%%%%%%%%%%%%%%%%%%%%%%%%%%%%%%%%%%%%%%%%%%%%%%%

\section{Three- and four-gauge-boson couplings}
\label{sec-tgcs}
\setcounter{equation}{0}

We now turn to the bounds on the anomalous couplings $h_i$ from measurements
of \mbox{$\gamma WW$} and \mbox{$ZWW$}~couplings at LEP2 \cite{Group:2002mc}
and the prospects to measure these couplings at a future~LC.  The former is
done in Sect.~\ref{ssec-tgcsatlep} using the scheme~$P_Z$, the latter in
Sect.~\ref{ssec-tgcsatnlc} using~$P_W$ and suitably defined effective~TGCs.  A
general parameterisation of the two triple-gauge-boson vertices by an
effective Lagrangian in the ELa approach (see Sect.~\ref{sec-intro}) requiring
only Lorentz invariance and Hermiticity consists of 14 real parameters.  A
common parameterisation used in the literature is the one of Hagiwara, Peccei,
Zeppenfeld and Hikasa~\cite{Hagiwara:1986vm}
\bea
\label{eq-lagrhagi}
\frac{{\mathscr L}^{\rm HPZH}_{VWW}}{ig_{VWW}} & = & g^V_1(W^+_{\mu
  \nu}W^{-\mu} - W^-_{\mu
  \nu}W^{+ \mu})V^{\nu}\\
& & \mbox{} + \kappa_V W_{\mu}^+ W^-_{\nu} V^{\mu \nu} +
  \frac{\lambda_V}{m_W^2} W_{\lambda \mu}^+ W_{\ \;\;\;\; \nu}^{-\mu} V^{\nu
  \lambda}\nonumber\\
& & \mbox{} + ig_4^V W_{\mu}^+ W^-_{\nu}
  (\partial^{\mu} V^{\nu} + \partial^{\nu} V^{\mu}) \nonumber \\
& & \mbox{} - i g_5^V \varepsilon^{\mu \nu \rho \sigma} \left(
  W_{\mu}^+ (\partial_{\rho} W^-_{\nu}) -
  W^-_{\nu} (\partial_{\rho} W_{\mu}^+)\right) V_{\sigma} \nonumber \\
& & \mbox{} +
  \tilde{\kappa}_V W_{\mu}^+ W^-_{\nu} \tilde{V}^{\mu \nu} +
  \frac{\tilde{\lambda}_V}{m_W^2} W_{\lambda \mu}^+ W_{\ \;\;\;\; \nu}^{- \mu}
  \tilde{V}^{\nu \lambda} \nonumber
\eea
with \mbox{$V = \gamma$} or~$Z$.  The overall constants for the photon and
$Z$~vertices are defined as follows:
\be
\label{eq-overall}
g_{\gamma WW} = - e,\msp \msp g_{ZWW} = - e \cot \theta_{\rm w},
\ee
where $e$ is the positron charge.  Then we have in the SM at tree-level
\be
\label{eq-smtgcs}
g_1^V = 1,\msp \msp \kappa_V = 1,
\ee
and all other couplings equal to zero.  We write $\Delta g_1^V = g_1^V - 1$
  and $\Delta \kappa_V = \kappa_V - 1$ as usual.  The $ZWW$~couplings involve
  the mixing angle~$\theta_{\rm w}$ of the SM.  In the ELa approach
  this~$\theta_{\rm w}$ is well defined.  It is also unique at least at
  tree-level.

Note that in the FF approach the same expression~(\ref{eq-lagrhagi}) is
usually written down but allowing the coupling constants to be complex
numbers.  Then ${\mathscr L}^{\rm HPZH}_{VWW}$ should not be considered as an
effective Lagrangian but only as a convenient shorthand description for the
\mbox{$VWW$} form factors generated by using~(\ref{eq-lagrhagi}) in Feynman
rules to first order.  In~\cite{Diehl:2002nj,Diehl:2003qz} the
parameterisation~(\ref{eq-lagrhagi}) is used and bounds on the anomalous
couplings are computed by means of optimal observables using the tree-level
expressions for the differential cross section of \mbox{$e^+ e^- \rightarrow
WW$}.  Given the expected accuracy at a future~LC it will in general be
necessary to take into account radiative corrections.  How this can be done in
the framework of optimal observables is explained in Sect.~3
of~\cite{Diehl:2002nj}.  One can apply to the measured cross section the~SM
radiative corrections in the reverse to obtain a Born-level cross section.
Neglecting again radiative corrections times and to anomalous couplings this
Born-level cross section can be analysed using tree graphs where for the~SM
(\ref{eq-smtgcs}) is valid.

Here we want to compare the parameters $h_i$ of our
Lagrangian~(\ref{eq-Leff})---which is in the ELb approach---to the parameters
in~(\ref{eq-lagrhagi}).  From the outset we must make it clear that such a
comparison raises problems.  In the ELa approach the dimension~\mbox{$\leq 4$}
terms in the Lagrangian are exactly the SM ones.  In the ELb approach
investigated in the present paper on the other hand the dimension~\mbox{$\leq
4$} terms receive anomalous contributions.  The relations between the~$h_i$
and the couplings $g_1^V$,\ldots, $\tilde{\lambda}_V$ of~(\ref{eq-lagrhagi})
which we shall derive below are thus only valid supposing that the anomalous
contributions to dimension~\mbox{$\leq 4$} terms are negligible.  For a
specific process one can take into account these contributions by defining
effective TGCs, as we shall do in Sect.~\ref{ssec-tgcsatnlc} below for the
reaction~\mbox{$e^+ e^- \rightarrow WW$}.

We now derive the relations of the parameters of~(\ref{eq-lagrhagi}) to the
  $h_i$ in the approximation where terms of the Lagrangian~(\ref{eq-Leff})
  that are of second or higher order in~$h_i$ are neglected.  The sine of the
  angle~$\theta_{\rm w}$ in~(\ref{eq-overall}) will be identified with $s_0$
  in the $P_Z$~scheme and with $s_1$ in the $P_W$~scheme.  The fact that we
  have an ambiguity here reflects again the differences of the ELa and ELb
  approaches.

We denote by~$\mathscr{L}_{\gamma WW}$ and $\mathscr{L}_{ZWW}$ the parts of
the Lagrangian~(\ref{eq-Leff})---expressed in terms of the physical fields
$W^{\pm}_{\mu}$, $A_{\mu}$ and $Z_{\mu}$---that consist of two $W$~boson
fields and one photon or $Z$-boson field, respectively.  Without any
approximation the \mbox{$\gamma WW$}~part is given by
\bea
\frac{\mathscr{L}_{\gamma WW}}{(-i e)} & = & \left(W^+_{\mu \nu} W^{-\mu} -
  W^-_{\mu \nu} W^{+\mu}\right) A^{\nu}\\
&& \mbox{} + \left(1 + \frac{c'_{\rm w}}{s'_{\rm
w}} \frac{h_{W \mskip -3mu B}}{\left(1 - h_{\varphi W}\right)}\right)
W^+_{\mu} W^-_{\nu} A^{\mu \nu}\nonumber\\
&& \mbox{} + \frac{6 \sqrt{2} G_{\rm F} s'_{\rm w}}{e \sqrt{d}} \frac{(1 +
  h_{\varphi}^{(1)}/2)}{\left(1 - h_{\varphi W}\right)} W^+_{\lambda \mu} W^{-
  \mu}_{\;\;\;\;\;\;\nu}\nonumber\\
&& \mbox{} \;\;\;\;\; \times \left(h_W A^{\nu \lambda} + h_{\tilde{W}}
  \tilde{A}^{\nu \lambda}\right)\nonumber\\
\label{eq-tgcsgamma}
&& {}+ \frac{c'_{\rm w}}{s'_{\rm w}} \frac{h_{\tilde{W} \mskip -3mu
  B}}{\left(1 - h_{\varphi W}\right)} W_{\mu}^+ W_{\nu}^- \tilde{A}^{\mu
  \nu},\nonumber
\eea
where \mbox{$\tilde{A}_{\mu \nu} = (1/2) \epsilon_{\mu \nu \rho \sigma}
  A^{\rho \sigma}$}, and $d$ is defined in~(\ref{eq-ddef}).  To obtain the
  term proportional to~$h_{\tilde{W}}$ in~(\ref{eq-tgcsgamma}) we have used
  the Shouten identity.  Depending on whether we are in the scheme $P_Z$ or
  $P_W$, $s'_{\rm w}$ is a solution to~(\ref{eq-swexact}) or~(\ref{eq-sww}),
  respectively.  The $ZWW$~part reads
\bea
\label{eq-tgcsz}
\frac{\mathscr{L}_{ZWW}}{(-i e)} & = & f_- \left(W^+_{\mu \nu}
W^{-\mu} - W^-_{\mu \nu} W^{+\mu}\right) Z^{\nu}\\
&& {} + \left(f_- - f_+ \frac{h_{W \mskip -3mu B}}{1 - h_{\varphi W}} \right)
W^+_{\mu} W^-_{\nu} Z^{\mu \nu}\nonumber\\
&& {}+ \hat{f} \frac{(1 + h_{\varphi}^{(1)}/2)}{\left(1 -
    h_{\varphi W}\right)} W^+_{\lambda \mu} W^{-
    \mu}_{\;\;\;\;\;\;\nu}\nonumber\\
&& {} \;\;\;\;\; \times \left(h_W Z^{\nu \lambda} + h_{\tilde{W}}
    \tilde{Z}^{\nu \lambda}\right)\nonumber\\
&& {} - f_+ \frac{h_{\tilde{W} \mskip -3mu B}}{1 - h_{\varphi W}} W_{\mu}^+
    W_{\nu}^- \tilde{Z}^{\mu \nu}\nonumber,
\eea
where \mbox{$\tilde{Z}_{\mu \nu} = (1/2) \epsilon_{\mu \nu \rho \sigma}
  Z^{\rho \sigma}$} and
\bea
f_+ & = & \frac{1}{\sqrt{t}} \left(d + \frac{b c'_{\rm w}}{s'_{\rm
w}}\right),\;\;\; f_- = \frac{1}{\sqrt{t}} \left(\frac{d
c'_{\rm w}}{s'_{\rm w}} - b\right),\\
\hat{f} & = & \frac{6 \sqrt{2} G_{\rm F} s'_{\rm w}}{e \sqrt{d}} f_-.
\eea
Again, for the term in~(\ref{eq-tgcsz}) proportional to~$h_{\tilde{W}}$ the
Shouten identity is applied.  Expanding the coefficients of the operators
in~(\ref{eq-tgcsgamma}) and~(\ref{eq-tgcsz}) to first order in the anomalous
couplings and comparing with the Lagrangian~(\ref{eq-lagrhagi}) we find the
following relations between the two sets of couplings, in the $P_Z$~scheme:
\begin{alignat}{3}
\label{eq-coup1}
\Delta g_1^{\gamma} & = 0,& \!\!\! \Delta \kappa_{\gamma} & = &&
\;\frac{c_0}{s_0} h_{W \mskip -3mu B},\\[.1cm]
\label{eq-coup2}
\Delta g_1^Z & = \frac{s_0}{c_0 \left(s_0^2 - c_0^2 \right)} h_{W
\mskip -3mu B} & \!\!\! \Delta \kappa_Z  & = && \frac{2 s_0 c_0}{s_0^2 -
c_0^2} h_{W \mskip -3mu B}\\
& + \frac{h_{\varphi}^{(3)}}{4 \left(s_0^2 - c_0^2\right)}, & &&& +
\frac{h_{\varphi}^{(3)}}{4 \left(s_0^2 - c_0^2\right)},\nonumber\\[.1cm]
\label{eq-coup3}
\lambda_Z & = 6 s_0 c_0^2 \sqrt{2} G_{\rm F} m_Z^2 h_W/e,&
\!\!\! \lambda_{\gamma} & = &&\; 6 s_0 c_0^2 \sqrt{2} G_{\rm F} m_Z^2
h_W/e,\\[.1cm] 
\label{eq-coup4}
\tilde{\kappa}_Z & = - \frac{s_0}{c_0} h_{\tilde{W} \mskip -3mu B},&
\!\!\! \tilde{\kappa}_{\gamma} & = &&\; \frac{c_0}{s_0} h_{\tilde{W} \mskip
-3mu B},\\[.1cm]
\label{eq-coup5}
\tilde{\lambda}_Z & = 6 s_0 c_0^2 \sqrt{2} G_{\rm F} m_Z^2 h_{\tilde{W}}/e,&
\!\!\! \tilde{\lambda}_{\gamma} & = &&\; 6 s_0 c_0^2 \sqrt{2} G_{\rm F} m_Z^2
h_{\tilde{W}}/e,\\[.1cm]
\label{eq-coup6}
g_4^{\gamma} & = g_4^Z = g_5^{\gamma} = g_5^Z = 0.&&&&
\end{alignat}
Equations~(\ref{eq-coup1}) to~(\ref{eq-coup3}) relate $CP$~conserving
couplings whereas (\ref{eq-coup4}) and~(\ref{eq-coup5}) relate $CP$~violating
ones.  The couplings $g_4^{\gamma}$ and $g_4^Z$ are $CP$~violating whereas
$g_5^{\gamma}$ and $g_5^Z$ are $CP$~conserving.  From~(\ref{eq-coup1})
to~(\ref{eq-coup6}) we see that in our ELb framework the anomalous
\mbox{$\gamma WW$} and \mbox{$ZWW$}~vertices depend only on five anomalous
parameters, three of them $CP$ conserving \mbox{($h_W$, $h_{W \mskip -3mu B}$,
$h_{\varphi}^{(3)}$)}, two of them $CP$ violating \mbox{($h_{\tilde{W}}$,
$h_{\tilde{W} \mskip -3mu B}$)}.  The 14 anomalous couplings
in~(\ref{eq-lagrhagi}) thus obey 9 relations.  These well known gauge
relations are
\bea
\label{eq-gr1}
\Delta g_1^{\gamma} & = & 0,\\
\label{eq-gr2}
\Delta \kappa_Z & = &\Delta g_1^Z - \frac{s_0^2}{c_0^2} \Delta
\kappa_{\gamma},\\
\label{eq-gr3}
\lambda_Z & = & \lambda_{\gamma},\\
\label{eq-gr4}
\tilde{\kappa}_{\gamma} & = & - \frac{c_0^2}{s_0^2}
\tilde{\kappa}_Z,\\
\label{eq-gr5}
\tilde{\lambda}_{\gamma} & = & \tilde{\lambda}_Z,\\
\label{eq-gr6}
g_4^{\gamma} & = & g_4^Z \;\,=\;\, g_5^{\gamma} \;\,=\;\, g_5^Z
\;\,=\;\, 0.
\eea
However, one has to keep in mind that although the number of TGCs is reduced
in the ELb approach compared to the ELa approach anomalous effects can occur
at other vertices or propagators, see e.g.\ our treatment of the reaction
\mbox{$e^+e^- \rightarrow WW$} in Sect.~\ref{ssec-tgcsatnlc}.  Notice also
that the gauge relations~(\ref{eq-gr1}) to~(\ref{eq-gr6}) do not generally
hold in an \mbox{$SU(2) \times U(1)$} invariant effective theory, but rather
stem from the fact that we have restricted ourselves to operators of
dimension~\mbox{$\leq 6$}.  If one adds to the Lagrangian~(\ref{eq-Leff})
suitable operators of higher dimension one can obtain a gauge-invariant
Lagrangian where all 14 anomalous couplings in (\ref{eq-lagrhagi}) are independent.  For this,
operators up to dimension~12 are required~\cite{Gounaris:1996rz}, where for
each additional dimension the effects are suppressed by an additional
factor~\mbox{$(v/\Lambda)$}.  The so-called gauge relations~(\ref{eq-gr1})
to~(\ref{eq-gr6}) are thus rather a low-energy approximation than a result
from gauge invariance.

Using the scheme~$P_W$, we find in the linear approximation instead
of~(\ref{eq-coup1}) to~(\ref{eq-coup6})
\begin{alignat}{3}
\label{eq-coup1a}
\Delta g_1^{\gamma} & = 0,&\Delta \kappa_{\gamma} & = &&\;\frac{c_1}{s_1} h_{W
  \mskip -3mu B},\\[.1cm]
\label{eq-coup2a}
\Delta g_1^Z & = 0,& \Delta \kappa_Z  & = &&\; - \frac{s_1}{c_1} h_{W
  \mskip -3mu B},\\[.1cm]
\label{eq-coup3a}
\lambda_Z & = 6 s_1 \sqrt{2} G_{\rm F} m_W^2 h_W / e,&
\lambda_{\gamma} & = &&\; 6 s_1 \sqrt{2} G_{\rm F} m_W^2 h_W / e,\\[.1cm]
\label{eq-coup4a}
\tilde{\kappa}_Z & = - \frac{s_1}{c_1} h_{\tilde{W} \mskip -3mu B},&
\tilde{\kappa}_{\gamma} & = &&\; \frac{c_1}{s_1} h_{\tilde{W} \mskip
-3mu B},\\[.1cm]
\label{eq-coup5a}
\tilde{\lambda}_Z & = 6 s_1 \sqrt{2} G_{\rm F} m_W^2 h_{\tilde{W}} / e,&
\tilde{\lambda}_{\gamma} & = &&\; 6 s_1 \sqrt{2} G_{\rm F} m_W^2 h_{\tilde{W}}
  / e,\\[.1cm]
\label{eq-coup6a}
g_4^{\gamma} & = g_4^Z = g_5^{\gamma} = g_5^Z = 0.&&&&
\end{alignat}
Notice that $h_{\varphi}^{(3)}$ does not enter here in~$P_W$ such that the
  number of couplings to describe the anomalous \mbox{$\gamma WW$} and
  \mbox{$ZWW$}~vertices in the $P_W$~scheme is one less than in the
  $P_Z$~scheme.  We have here two $CP$~conserving couplings \mbox{($h_W$,
  $h_{W \mskip -3mu B}$}) and two $CP$~violating ones \mbox{($h_{\tilde{W}}$,
  $h_{\tilde{W} \mskip -3mu B}$}).  The gauge relations~(\ref{eq-gr1})
  to~(\ref{eq-gr6}) also hold in the scheme~$P_W$ if we substitute $s_0$ and
  $c_0$ by $s_1$ and~$c_1$.  In the $P_W$~scheme we have a further gauge
  relation
\be
\label{eq-grg1z}
\Delta g_1^Z = 0.
\ee
Thus we find in our locally \mbox{$SU(2) \times U(1)$}~symmetric theory that
the number of independent $CP$~conserving TGCs is three if we choose the
$P_Z$~scheme.  This agrees with the results of~\cite{Grosse-Knetter:1993rp}.
If we choose~$P_W$, which is actually the convenient scheme for the direct
measurement of TGCs in $W$-boson-pair production there is one TGC less.
However, the $h_i$ also enter in fermion-boson vertices, Higgs-boson vertices
and boson masses.  In fact, we shall see in Sect.~\ref{ssec-tgcsatnlc} that
the coupling~$h_{\varphi}^{(3)}$ affects the differential cross section of
\mbox{$e^+ e^- \rightarrow WW$} although we use the scheme~$P_W$.

Without approximation the \mbox{$\gamma \gamma WW$} part of~(\ref{eq-Leff}) is
\bea
\label{eq-l4}
\frac{\mathcal{L}_{\gamma \gamma WW}}{(-e^2)} & = & \left(W^+_{\mu} W^{- \mu}
  A_{\nu} A^{\nu} - W^+_{\mu} W^-_{\nu} A^{\mu} A^{\nu}\right)\\
 && {} - \frac{6 s'_{\rm w}}{e v^2 \sqrt{d}} \frac{h_W A_{\lambda
  \mu} + h_{\tilde{W}} \tilde{A}_{\lambda
  \mu}}{(1 - h_{\varphi W})}\nonumber\\
& & \mbox{} \;\;\; \times \Big(\left(A^{\mu} W^+_{\nu} - A_{\nu} W^{+
  \mu}\right) W^{- \nu \lambda} + {\rm H.c.}\Big). \nonumber
\eea
Using the formulae of Sect.~\ref{sec-para} it is straightforward to calculate
the linear approximation of~(\ref{eq-l4}) for the two schemes.

The terms containing two photon fields and one Higgs field in the effective
Lagrangian~(\ref{eq-Leff}) after diagonalisation are, without approximation,
\bea
v d \; \sqrt{1 + \big(h_{\varphi}^{(1)} +
    h_{\varphi}^{(3)}\big)\! /2} \;\;\mathscr{L}_{\gamma\gamma H} =
  \;\;\;\;\;\;\;\;\;\;\;\;\;\;\;\;\;\;\;\;\;\;\;\;\;\;\;\;\; && \\
\frac{1}{2}
\left(s_{\rm w}^{\prime \, 2} h_{\varphi W} + c_{\rm w}^{\prime \,
2} h_{\varphi B} - 2 c_{\rm w}^{\prime} s_{\rm w}^{\prime} h_{W \mskip
-3mu B}\right) A_{\mu\nu} A^{\mu\nu} H && \nonumber\\
 + \left(s_{\rm w}^{\prime \, 2} h_{\varphi \tilde{W}} +
  c_{\rm w}^{\prime \, 2} h_{\varphi \tilde{B}} - c_{\rm w}^{\prime} s_{\rm
w}^{\prime} h_{\tilde{W} \mskip -3mu B}\right)  \tilde{A}_{\mu \nu}
A^{\mu \nu} H.&&\nonumber
\eea
In the linear approximation we simply have to drop the square root, and
substitute the factor~\mbox{$vd$} on the left hand side by \mbox{$(\sqrt{2}
G_{\rm F})^{-1/2}$} and $s_{\rm w}^{\prime}$ ($c_{\rm w}^{\prime}$) on the
right hand side by $s_0$ ($c_0$) in the $P_Z$~scheme, and by $s_1$ ($c_1$) in
the $P_W$~scheme.

We summarise in Tab.~\ref{tab:vertices} which couplings contribute to the
\mbox{$\gamma WW$}, \mbox{$ZWW$}, \mbox{$\gamma \gamma WW$} and \mbox{$\gamma
\gamma H$}~vertices if we consider only terms that are linear in the~$h_i$.
\begin{table*}
\centering
\begin{tabular}{rccccccccccc}
\hline
 &&&&&&&&&&&\\[-.28cm]
 & SM & $h_W$ & $h_{\tilde{W}}$ & $h_{\varphi W}$ & $h_{\varphi \tilde{W}}$
 & $h_{\varphi B}$ & $h_{\varphi \tilde{B}}$ & $h_{W \mskip -3mu B}$ &
 $h_{\tilde{W} \mskip -3mu B}$ & $h_{\varphi}^{(1)}$ &
 $h_{\varphi}^{(3)}$\\[.06cm]
\hline
 &&&&&&&&&&&\\[-.25cm]
$\gamma WW$ & $\surd$ & $\surd$ & $\surd$ & & & & &
 $\surd$ & $\surd$ & &\\
$ZWW$ & $\surd$ & $\surd$ & $\surd$ & & & & & $\surd$ & $\surd$ &  & $P_Z$\\
$\gamma \gamma WW$ & $\surd$ & $\surd$ & $\surd$ & & & & & & & &\\
$\gamma \gamma H$ & & & & $\surd$ & $\surd$ & $\surd$ & $\surd$ & $\surd$ &
 $\surd$ & &\\[.05cm]
\hline
\end{tabular}
\caption{\label{tab:vertices}Contributions of the SM Lagrangian and of
the anomalous operators to different vertices in linear order in the~$h_i$
after the simultaneous diagonalisation.  Only those vertices are listed that
are relevant for our observables.  This does not coincide with the
contributions to operators of the respective structure before the simultaneous
diagonalisation, see Tab.~\protect{\ref{tab:contri}}.  The
coupling~$h_{\varphi}^{(3)}$ contributes to the $ZWW$~vertex in the
scheme~$P_Z$ but not in~$P_W$.}
\end{table*}
%
%

%%%%%%%%%%%%%%%%%%%%%%%%%%%%%%%%%%%%%%%%%%%%%%%%%%%%%%%%%%%%%%%%%

\subsection{Bounds from LEP2}
\label{ssec-tgcsatlep}

\begin{figure*}
\begin{center}
\includegraphics[totalheight=4.5cm]{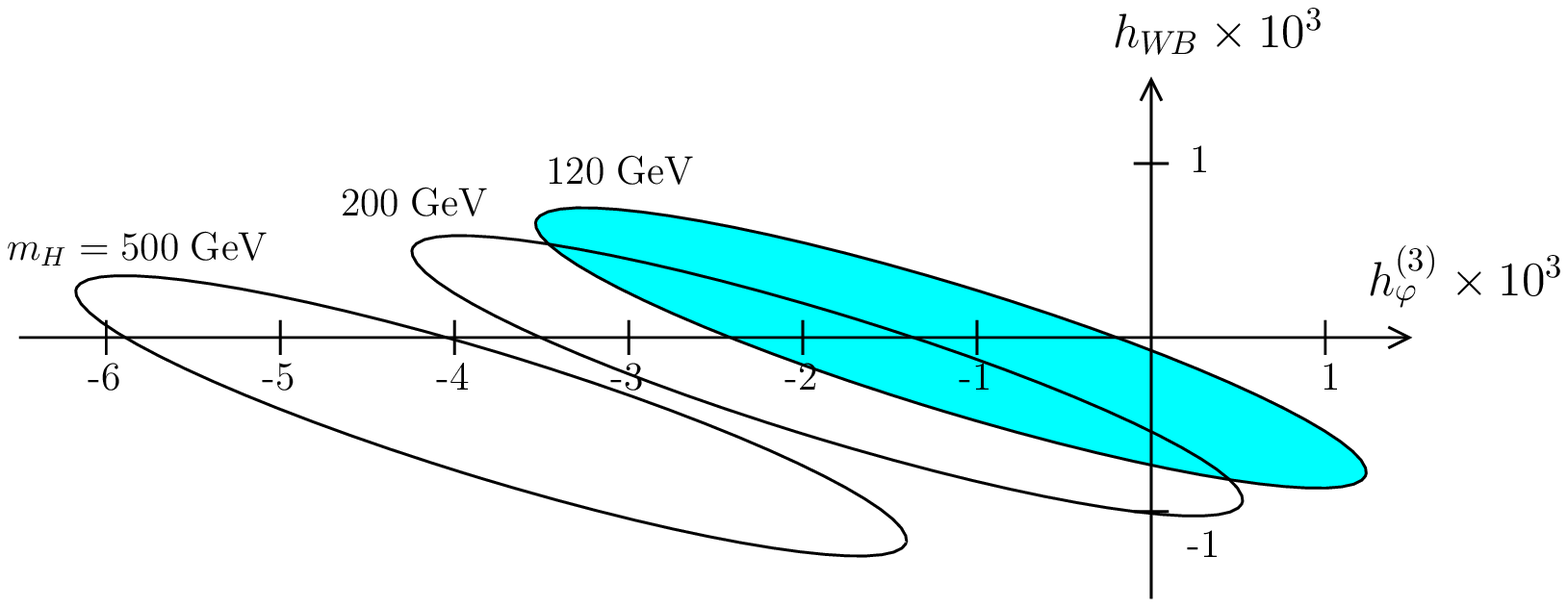}
\end{center}
\caption{\label{fig:precdata}Error ellipses of~$h_{W \mskip -3mu B}$
  and~$h_{\varphi}^{(3)}$ for different Higgs masses.}
\end{figure*}

For the $CP$ conserving couplings we use the values from Tab.~11.7
in~\cite{Group:2002mc}
\begin{alignat}{2}
\label{eq-val}
\Delta g_1^Z & = &\;0.051 &\pm 0.032,\\
\Delta \kappa_{\gamma} & = &\;-0.067 &\pm 0.061,\nonumber\\
\lambda_{\gamma} & = &\;-0.067 &\pm 0.038.\nonumber
\end{alignat}
The errors given in~\cite{Group:2002mc} are not symmetric.  Here we make the
conservative choice to take the larger of the lower and upper errors.  The
correlations, in the order \mbox{$\Delta g_1^Z$}, \mbox{$\Delta
\kappa_{\gamma}$}, $\lambda_{\gamma}$ from the same reference are
\be
\label{eq-corrcoup}
\left(\ba{rrr}     
1 & 0.23 & -0.30\\
 & 1 & -0.27\\
 && 1
\ea\right).
\ee
The remaining two non-zero $CP$~conserving couplings \mbox{$\Delta \kappa_Z$}
and \mbox{$\lambda_Z$} are not considered as independent
in~\cite{Group:2002mc}, but are assumed to be given by the gauge
relations~(\ref{eq-gr2}) and~(\ref{eq-gr3}).  From the values~(\ref{eq-val})
and~(\ref{eq-corrcoup}) we therefore obtain, using~(\ref{eq-coup1})
to~(\ref{eq-coup3}), the following values and errors for our anomalous
couplings
\begin{alignat}{2}
\label{eq-coupres}
h_W & = &\; -0.069 &\pm 0.039,\\
h_{W \mskip -3mu B} & = &\; -0.037 &\pm 0.033,\nonumber\\
h_{\varphi}^{(3)} & = &\; -0.029 &\pm 0.112,\nonumber
\end{alignat}
and the correlations, in the order $h_W$, $h_{W \mskip -3mu B}$,
$h_{\varphi}^{(3)}$,
\be
\label{eq-tgcscorr}
\left(\ba{rrr}     
1 & -0.27 & 0.36\\
 & 1 & -0.80\\
 && 1
\ea\right).
\ee
We repeat that these constraints are only approximate as in our ELb framework
non-SM effects do not only occur at the three-boson vertices, but also at the
fermion-boson vertices and through~$m_W$.  The bounds~(\ref{eq-coupres}) on
the~$h_i$ are thus only valid to the approximation that these effects are
negligible.\footnote{In the following subsection we show that one can take
into account the effects from anomalous fermion-boson couplings and anomalous
boson masses by defining effective~TGCs.  However, to this end each physics
reaction must be considered separately.  Here we use the combined results from
various processes and one cannot easily avoid this simplification.}  Moreover,
in contrast to Sect.~\ref{sec-lep}, no radiative corrections are included in
our results here.  The constraints on $h_{W \mskip -3mu B}$ and
$h_{\varphi}^{(3)}$ derived from TGC measurements are much weaker than the
constraints from Tab.~\ref{tab:coup2}.  Combining the results from
Tab.~\ref{tab:coup2} with~(\ref{eq-coupres}) and~(\ref{eq-tgcscorr}) we find
the values and errors as listed in Tab.~\ref{tab:coup3}.  These are the final
values for the $CP$~conserving couplings that we can derive from LEP1, SLC,
LEP2 and $W$-boson measurements.  The value and error of~$h_W$ is almost
independent of~$m_H$.  Electroweak data predicts a value for~$h_W$ of
about~$-0.06$.  Since the errors on~$h_{W \mskip -3mu B}$
and~$h_{\varphi}^{(3)}$ are almost uncorrelated with the error on~$h_W$, we
can consider the bounds on~$h_{W \mskip -3mu B}$ and~$h_{\varphi}^{(3)}$
separately.  Their error ellipses are shown in Fig.~\ref{fig:precdata}.
Interestingly, a large Higgs mass is allowed by the data if~$h_{W \mskip -3mu
B}$ and~$h_{\varphi}^{(3)}$ are of order~\mbox{$\sim 10^{-3}$}.
\begin{table*}
\centering
\begin{tabular}{lrrrrrrrr}
\hline
\hline
 &&&&&&&&\\[-.31cm]
 && \multicolumn{4}{l}{$s_{\rm eff}^2$, $\Gamma_Z$, $\sigma^0_{\rm had}$,
 $R^0_{\ell}$, $m_W$, $\Gamma_W$, TGCs} &&&\\
\hline
\hline
 &&&&&&&&\\[-.34cm]
$m_H$& \hspace{-.5cm} & 120 GeV & 200 GeV & 500 GeV & $\delta h \times 10^3$
&&&\\
\hline
 &&&&&&&&\\[-.32cm]
$h_W$ \hspace{-.5cm} & $\times 10^3$ & $-62.4$ & $-62.5$ & $-62.8$ & 36.3 &
\msp 1 & $-0.007$ & $0.008$\\
$h_{W \mskip -3mu B}$ \hspace{-.5cm} & $\times 10^3$ & $-0.06$ & $-0.22$ &
$-0.45$ & 0.79 & & 1 & $-0.88$\\
$h_{\varphi}^{(3)}$ \hspace{-.5cm} & $\times 10^3$ & $-1.15$ & $-1.86$ &
$-3.79$ & 2.39 & & & 1\\[.02cm]
\hline
\hline
\end{tabular}
\caption{\label{tab:coup3}Final results from already existing data for
  $CP$~conserving couplings in units of $10^{-3}$ for a Higgs mass of 120~GeV,
  200~GeV and 500~GeV.  The anomalous couplings are extracted from the
  observables listed in the first row using~(\protect{\ref{eq-seffres}}).  The
  errors \mbox{$\delta h$} and the correlations of the errors are independent
  of the Higgs mass with the accuracy given here.  The correlation matrix is
  given on the right.}
\end{table*}
\begin{table}
\centering
\begin{tabular}{lrr}
\hline
\hline
 &&\\[-.32cm]
 & \multicolumn{2}{c}{TGCs}\\
\hline
\hline
&&\\[-.25cm]
& $h$ & $\delta h$\\
\hline
&&\\[-.32cm]
$h_{\tilde{W}}$ & 0.068 & 0.081\\
$h_{\tilde{W} \mskip -3mu B}$ & 0.033 & 0.084\\[.05cm]
\hline
\hline
\end{tabular}
\caption{\label{tab:coup4}Final results from already existing data for
  $CP$~violating couplings.  The anomalous couplings are extracted from TGC
  measurements at LEP2 in various processes.}
\end{table}

For the $CP$~violating couplings we use the weighted average of the single
parameter measurements given in \cite{Abbiendi:2000ei}
and~\cite{Heister:2001qt}
\be
\label{eq-val4}
\tilde{\lambda}_Z = 0.067 \pm 0.080,\;\;\;\;\;\;\;\;\;\;
\tilde{\kappa}_Z = -0.018 \pm 0.046.
\ee
In these analyses the relations~(\ref{eq-gr4}) and~(\ref{eq-gr5}) of the
$CP$~violating photon couplings with the $CP$~violating $Z$~couplings are
assumed to hold.  Using the values~(\ref{eq-val4}) we get
from~(\ref{eq-coup4}) and~(\ref{eq-coup5}) the results listed in
Tab.~\ref{tab:coup4}.  These results are independent of~$m_H$.  Since---in
contrast to the $CP$~conserving couplings---the $CP$~violating couplings do
not affect the boson-fermion couplings or the $W$~mass these bounds are
accurate in the sense that no such effects are neglected.

Bounds at 95\%~C.L.\ on anomalous TGCs have been determined by the
CDF~collaboration~\cite{Abe:1994fx} and the
\mbox{DO\hspace{-.23cm}\slash}\hspace{.15cm}collaboration~\cite{Abbott:1999ec}.
The latter, who gives the tighter constraints, also quotes central values and
68\%~C.L.\ limits on $\lambda_{\gamma}$ and \mbox{$\Delta \kappa_{\gamma}$}.
They are \mbox{$\lambda_{\gamma} = 0.00^{+0.10}_{-0.09}$} and \mbox{$\Delta
\kappa_{\gamma} = -0.08^{+0.34}_{-0.34}$}, and therefore not tighter than the
constraints~(\ref{eq-val}) from LEP2.  Moreover, the values~(\ref{eq-val}) are
results where all three parameters are measured at a time.
In~\cite{Abe:1994fx} also 95\%~C.L.~limits on two $CP$~violating couplings are
determined, viz.\ \mbox{$-0.7 < \tilde{\lambda}_{\gamma} < 0.7$} and
\mbox{$-2.3 < \tilde{\kappa}_{\gamma} < 2.2$}.  These results can be
transformed using~(\ref{eq-gr4}) and~(\ref{eq-gr5}) into bounds on the
couplings $\tilde{\lambda}_Z$ and \mbox{$\tilde{\kappa}_Z$} at 68\%~C.L.
These resulting bounds are less stringent than the LEP2~bounds
(\ref{eq-val4}).  We thus conclude that an inclusion of the bounds
from~\cite{Abe:1994fx,Abbott:1999ec} would not have a considerable effect on
our calculated bounds on the~$h_i$.

As mentioned above, see~(\ref{eq-order}), a natural choice for the
coefficients $h_i$ in (\ref{eq-Leff2}) is \mbox{$h_i = \alpha_i v^2/
\Lambda^2$} where $\Lambda$ is the new-physics scale and the $\alpha_i$ are of
order one.  Setting \mbox{$\alpha_i = 1$} and using the numbers from
Tabs.~\ref{tab:coup3} and~\ref{tab:coup4} we find lower bounds~$\Lambda_i$ on
the scale of new physics according to
\be
\label{eq-bounds}
\Lambda_i = \frac{v}{\sqrt{|h_i| + \delta h_i}}.
\ee
These bounds are listed in Tab.~\ref{tab:scale}.  New physics that gives rise
to non-zero $h_W$, $h_{\tilde{W}}$ or $h_{\tilde{W} \mskip -3mu B}$ may be
seen at a LC in the one-TeV-range.  Those affecting $h_{\varphi}^{(3)}$ can
lead to visible effects at a multi-TeV machine like CLIC, whereas $h_{W \mskip
-3mu B}$ will probably be out of reach in the near future.  We remark that
relations between the Higgs mass and the scale of new physics in an
effective-Lagrangian approach have also been obtained using renormalisation
group methods, see~\cite{Grzadkowski:2003sd}.  There operators of dimension
six containing the Higgs and the top-quark fields are included in the
effective Lagrangian, and triviality and vacuum-stability arguments are
applied.
\begin{table}
\centering
\begin{tabular}{lrrr}
\hline
 &&&\\[-.32cm]
$m_H$ [GeV] & 120 & 200 & 500\\
\hline
 &&&\\[-.32cm]
$h_W$ & 0.78 & 0.78 & 0.78\\
$h_{W \mskip -3mu B}$ & 8.4 & 7.7 & 7.0\\
$h_{\varphi}^{(3)}$ & 4.1 & 3.8 & 3.1\\
$h_{\tilde{W}}$ & 0.64 & 0.64 & 0.64\\
$h_{\tilde{W} \mskip -3mu B}$ & 0.72 & 0.72 & 0.72\\[.04cm] 
\hline
\end{tabular}
\caption{\label{tab:scale}Lower bounds $\Lambda_i$ on the new-physics scale
  $\Lambda$ in TeV from the values of different anomalous couplings $h_i$
  obtained from the results in Tabs.~\ref{tab:coup3} and~\ref{tab:coup4}
  according to~(\ref{eq-bounds}).  The numbers are given for a Higgs mass of
  120~GeV, 200~GeV and 500~GeV, respectively.}
\end{table}

To first order in the anomalous couplings none of the observables considered
so far depends on $h_{\varphi W}$, $h_{\varphi \tilde{W}}$, $h_{\varphi B}$,
$h_{\varphi \tilde{B}}$ or $h_{\varphi}^{(1)}$.  This does not change when
taking into account optimal observables for \mbox{$e^+ e^- \rightarrow WW$}
with the effective couplings, see Sect.~\ref{ssec-tgcsatnlc}.  However, four
couplings that cannot be determined with present data or in \mbox{$e^+ e^-
\rightarrow WW$} at a future LC have an impact on the differential cross
section for $W$-pair production at a photon collider, which we will study in a
future work~\cite{photon-collider}.  To be precise, one linear combination of
$h_{\varphi W}$ and $h_{\varphi B}$ and one linear combination of $h_{\varphi
\tilde{W}}$ and $h_{\varphi \tilde{B}}$ can be measured including data from
this reaction.  Then only three anomalous-coupling combinations, that is the
other two linear combinations of these four couplings as well
as~$h_{\varphi}^{(1)}$, cannot be determined.  We summarise this result in
Tab.~\ref{tab:sum} where we show which coupling combinations can be measured
by means of which observables.  In the right column we list all observables
that we use in this work or in~\cite{photon-collider}.
\begin{table*}
\begin{alignat}{2}
\hline
\hline \nonumber
&&& \\[-.46cm]
& P_Z{\rm \ scheme} && \nonumber\\
\hline
\hline \nonumber
&&& \\[-.46cm]
& h_{W \mskip -3mu B},\, h_{\varphi}^{(3)} &&
s_{\rm eff}^2,\, \Gamma_Z,\, \sigma^0_{\rm had},\, R^0_{\ell},\, m_W,\,
\Gamma_W \nonumber\\
& h_W,\, h_{W \mskip -3mu B},\, h_{\varphi}^{(3)}
&& {\rm 3}~CP~{\rm conserving~TGCs} \nonumber\\
& h_{\tilde{W}},\, h_{\tilde{W} \mskip -3mu B} && {\rm
2}~CP~{\rm violating~TGCs} \nonumber\\[.14cm]
\hline
\hline \nonumber
&&& \\[-.46cm]
& P_W{\rm \ scheme} && \nonumber\\
\hline
\hline \nonumber
&&& \\[-.46cm]
& h_W,\, h_{W \mskip -3mu B},\, h_{\varphi}^{(3)},\, h_{\tilde{W}},\,
h_{\tilde{W} \mskip -3mu B} && {\rm effective~couplings~in}~e^+e^- \rightarrow
WW 
\nonumber\\
& \hspace{-.16cm}\begin{array}{l}
h_W,\, h_{W \mskip -3mu B},\, h_{\tilde{W}},\, h_{\tilde{W} \mskip
-3mu B},\\[.1cm]
(s_1^2 h_{\varphi W} + c_1^2 h_{\varphi B}),\, (s_1^2 h_{\varphi \tilde{W}} +
c_1^2 h_{\varphi \tilde{B}})
\end{array} \Bigg\} \;\;\;\; && {\rm optimal~observables~for}~\gamma \gamma
\rightarrow WW \nonumber\\[.11cm] 
\hline
\hline \nonumber
\end{alignat}
\caption{\label{tab:sum}Anomalous couplings and observables for their
  measurement in the respective schemes, in which they are considered in our
  studies.  With the ensemble of all these observables five couplings can be
  measured independently.  In addition, of the two couplings $h_{\varphi W}$
  and~$h_{\varphi B}$ one linear combination can be extracted.  The same is
  true for~$h_{\varphi \tilde{W}}$ and~$h_{\varphi \tilde{B}}$.}
\end{table*}

%%%%%%%%%%%%%%%%%%%%%%%%%%%%%%%%%%%%%%%%%%%%%%%%%%%%%%%%%%%%%%%%%

\subsection[Effective couplings for $e^+ e^- \rightarrow WW$]{Effective
  couplings for \boldm{e^+ e^- \rightarrow WW}}
\label{ssec-tgcsatnlc}

\begin{figure}
\centering
\includegraphics[totalheight=3cm]{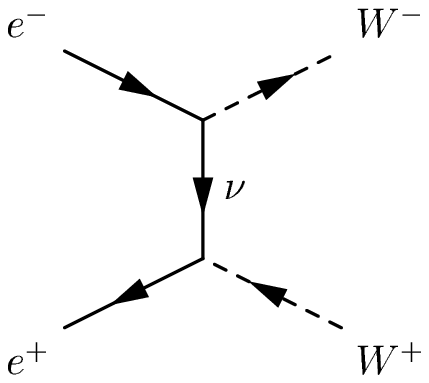}
\caption{\label{fig:neutrino}Neutrino-exchange diagram.}
\end{figure}

\begin{figure}
\begin{minipage}{0.24\textwidth}
\centering
\includegraphics[totalheight=3cm]{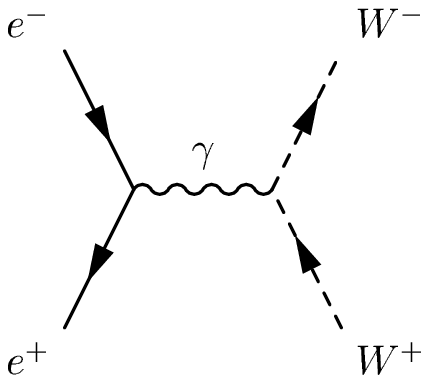}\\
(a)
\end{minipage}
\begin{minipage}{0.24\textwidth}
\centering
\includegraphics[totalheight=3cm]{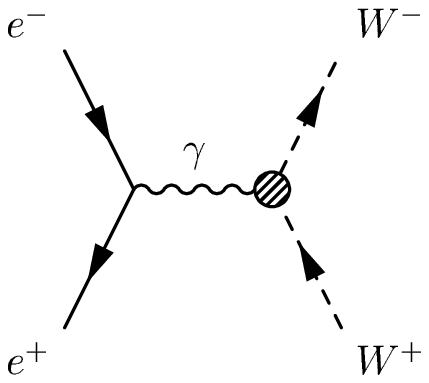}\\
(b)
\end{minipage}
\caption{\label{fig:photon}Photon-exchange diagrams.  SM diagram (a) and
  diagram with anomalous \mbox{$\gamma WW$} couplings~(b).}
\end{figure}

\begin{figure}
\begin{minipage}{0.24\textwidth}
\centering
\includegraphics[totalheight=3cm]{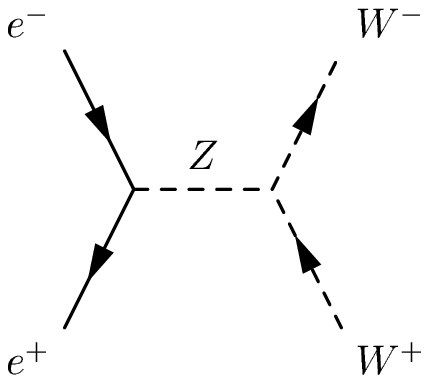}\\
(a)
\end{minipage}
\begin{minipage}{0.24\textwidth}
\centering
\includegraphics[totalheight=3cm]{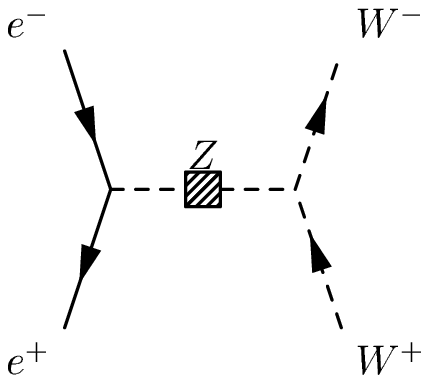}\\
(b)
\end{minipage}
\begin{minipage}{0.24\textwidth}
\centering
\includegraphics[totalheight=3cm]{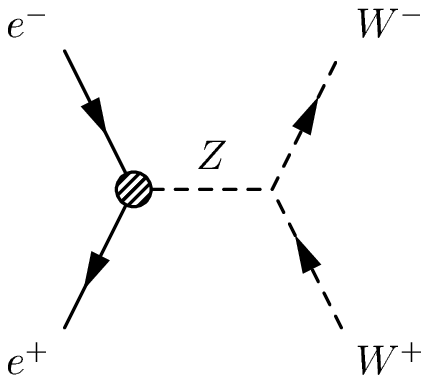}\\
(c)
\end{minipage}
\begin{minipage}{0.24\textwidth}
\centering \includegraphics[totalheight=3cm]{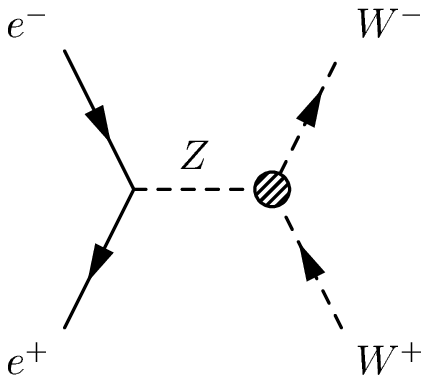}\\ 
(d)
\end{minipage}
\caption{\label{fig:zboson}$Z$-boson-exchange diagrams.  SM diagram~(a) and
  anomalous contributions from the modification of the $Z$~mass~(b), from
  anomalous \mbox{$eeZ$} couplings~(c) and anomalous
  \mbox{$ZWW$}~couplings~(d).}
\end{figure}

Here we would like to derive bounds on the anomalous couplings $h_i$ from
results obtained for the reaction \mbox{$e^+e^- \rightarrow WW$}
in~\cite{Diehl:2002nj,Diehl:2003qz}.  There all 14 complex parameters to
describe the general \mbox{$\gamma WW$} and \mbox{$ZWW$}~vertices are taken
into account, see~(\ref{eq-lagrhagi}), but the fermion-boson vertices, $m_Z$
and $m_W$ are supposed to be as in the~SM.  Therefore we have to analyse
carefully to which extent bounds on our anomalous couplings~$h_i$ can be
obtained from~\cite{Diehl:2002nj,Diehl:2003qz}.  Consider the two cases, the
ELb framework using the Lagrangian~(\ref{eq-Leff}) with all anomalous
couplings and the ELa framework of the Lagrangian~(\ref{eq-lagrhagi}) with
only anomalous~TGCs.  In both cases the process \mbox{$e^+e^- \rightarrow WW$}
has to be calculated at tree-level from three diagrams, \mbox{$t$-channel}
neutrino exchange, \mbox{$s$-channel} photon and \mbox{$s$-channel}
$Z$~exchange, see Figs.~\ref{fig:neutrino} to~\ref{fig:zboson}.  The various
anomalous contributions in each figure are explained below.  Given the
projected accuracy at a future~LC, it will in general be necessary to take
into account radiative corrections to the process~\mbox{$e^+ e^- \rightarrow
4$}~fermions within the~SM, which have been worked out in detail in the
literature~\cite{Grunewald:2000ju}.  How these corrections can be included in
an analysis with optimal observables is explained in Sect.~3
of~\cite{Diehl:2002nj}.  See also the discussion after~(\ref{eq-smtgcs})
above.  In~\cite{Diehl:2002nj,Diehl:2003qz} to linear order in the anomalous
TGCs the errors on their imaginary parts are not correlated with the errors on
their real parts.  This is because integrated observables are used and the
respective anomalous amplitudes obtain different signs under the combined
discrete symmetry \mbox{$CP\tilde{T}$} of $CP$ and a na\"{\i}ve time reversal
$\tilde{T}$, that is the simultaneous flip of all spins and momenta without
interchanging initial and final state.  Thus, whether or not the imaginary
parts are included in the analyses of \cite{Diehl:2002nj,Diehl:2003qz} plays
no r\^{o}le when we look at the sensitivity to the real parts.  For the real
parts, the errors on the $CP$~conserving couplings are not correlated with the
ones on the $CP$~violating couplings in the linear approximation, and the two
groups of couplings can be considered separately,
see~\cite{Diehl:2002nj,Diehl:2003qz}.  In principle, the derivation of bounds
on the~$h_i$ would require a complete calculation of the process \mbox{$e^+e^-
\rightarrow WW \rightarrow$}~4~fermions in the framework of the
Lagrangian~(\ref{eq-Leff}).  To first order in the couplings the errors on
$CP$~conserving and $CP$~violating couplings are not correlated also in this
case.  However, in such an analysis also anomalous effects from the couplings
of the $Z$~boson to fermions, which modify the \mbox{$s$-channel} $Z$~exchange
as well as anomalous contributions to $m_W$ ($m_Z$) must be taken into account
if we use the scheme~$P_Z$~($P_W$), see~(\ref{eq-wmass})
and~(\ref{eq-pwzmass}).  Furthermore, in the scheme $P_Z$ the anomalous
couplings have an impact on the couplings of the $W$~boson to fermions,
whereas in $P_W$ they have not due to~(\ref{eq-ccpw}).  As mentioned in the
introduction of Sect.~\ref{sec-para}, $m_W$ is treated as a fixed parameter
in~\cite{Diehl:2002nj,Diehl:2003qz}.  Thus for the analysis in this section it
is convenient to choose the $P_W$~scheme.  Moreover this simplifies the
analysis because in~$P_W$ the neutrino-exchange amplitude contains no
anomalous effects.  The $CP$~{\em violating} couplings appear in the reaction
\mbox{$e^+e^- \rightarrow WW$} only at the three-gauge-boson vertices.  Thus
the errors and correlations of these couplings can be obtained directly from
the results in~\cite{Diehl:2002nj,Diehl:2003qz} by using (\ref{eq-coup4a})
to~(\ref{eq-coup6a}).  In contrast, in the $CP$~{\em conserving} case we
obtain anomalous contributions to the vertices $eeZ$, \mbox{$\gamma WW$} and
$ZWW$ and to $m_Z$ from the Lagrangian~(\ref{eq-Leff}).  Therefore in the
framework of the Lagrangian~(\ref{eq-Leff}), all diagrams of
Figs.~\ref{fig:neutrino} to~\ref{fig:zboson} contribute to \mbox{$e^+e^-
\rightarrow WW$} in zeroth or linear order in the~$h_i$.  The blobs denote
anomalous couplings (without the SM contribution to the respective vertex) and
the diagram~(b) in Fig.~\ref{fig:zboson} with the box denotes
\mbox{$s$-channel} $Z$-boson exchange with a modified $Z$~mass in the
propagator {\em minus} the SM diagram, which is the diagram~(a).  Notice that
the $W$-decay amplitudes remain unchanged by the $h_i$ in the $P_W$~scheme.


After this discussion of the calculation of the amplitude for \mbox{$e^+ e^-
  \rightarrow WW$} in our present ELb approach we compare it to the FF
  calculation of~\cite{Diehl:2002nj,Diehl:2003qz} which can be considered as
  an ELa approach if we set all imaginary parts of coupling constants there to
  zero.  In the ELa framework of~\cite{Diehl:2002nj,Diehl:2003qz} the diagrams
  of Figs.~\ref{fig:neutrino} and~\ref{fig:photon} and only (a) and (d) of
  Fig.~\ref{fig:zboson} occur.  We will now show that the diagrams (b) and (c)
  of Fig.~\ref{fig:zboson}, that is the anomalous effects at the
  \mbox{$eeZ$}~coupling and in $m_Z$, can be completely shifted to diagram~(b)
  in Fig.~\ref{fig:photon} and diagram~(d) in Fig.~\ref{fig:zboson} by
  defining new {\it effective} \mbox{$\gamma WW$} and \mbox{$ZWW$}~couplings.
  For given values of the couplings~$h_i$, which modify the TGCs, the
  fermion-boson couplings and $m_Z$ in the ELb framework of the
  Lagrangian~(\ref{eq-Leff}), we can compute values for these effective
  anomalous~TGCs.  Then calculating the process \mbox{$e^+e^- \rightarrow WW$}
  in the ELa framework~(\ref{eq-lagrhagi}) of~\cite{Diehl:2002nj,Diehl:2003qz}
  with merely (effective) anomalous TGCs leads to the same differential cross
  section as calculating it with all anomalous vertices in~ELb.  This means
  the amplitudes for the process are only computed from the diagram in
  Fig.~\ref{fig:neutrino}, both diagrams in Fig.~\ref{fig:photon} and diagrams
  (a) and (d) in Fig.~\ref{fig:zboson}, but with suitably defined effective
  \mbox{$\gamma WW$} and \mbox{$ZWW$}~couplings.

We start from the Lagrangian~(\ref{eq-Leff}) and denote the parts of the
amplitudes for \mbox{$e^+ e^- \rightarrow WW$} obtained from the tree-level
diagrams for \mbox{$t$-channel} neutrino exchange, and \mbox{$s$-channel}
photon and $Z$~exchange by $\mathcal{A}_{\nu}$, $\mathcal{A}_{\gamma}$ and
$\mathcal{A}_{Z}$, respectively.  First we assume that these amplitudes are
the full expressions without linearisation in the~$h_i$.  Thus these
amplitudes do not correspond to the sum of the diagrams in
Figs.~\ref{fig:neutrino} to~\ref{fig:zboson}, where we have assumed that all
terms of second or higher order in the anomalous couplings are neglected and
the diagrams with the various anomalous contributions can therefore be summed
linearly.  The linearisation is done in a second step below.  The
amplitude~$\mathcal{A}_{\nu}$ is identical to the neutrino \mbox{$t$-channel}
exchange in the~SM.  The amplitude~$\mathcal{A}_{\gamma}$ is affected by the
anomalous couplings only at the \mbox{$\gamma WW$}~vertex.  However, we will
define effective \mbox{$\gamma WW$}~couplings below because some contributions
from the $Z$~exchange will be carried over to the photon exchange.  The
amplitude~$\mathcal{A}_Z$ is affected by anomalous couplings at the
\mbox{$eeZ$} and \mbox{$ZWW$} vertices, as well as through~$m_Z$.  Now
consider the currents (\ref{eq-defemc}) and~(\ref{eq-neutrc}) for a certain
charged lepton species~$\ell$ (in our case $\ell$ is the electron):
\bea
\mathcal{J}^{\mu}_{\rm em} (\ell) & = & \overline{\ell} \gamma^{\mu}
(\mathbf{T}_3 + \mathbf{Y}) \ell, \\
\mathcal{J}^{\mu}_{\rm NC} (\ell) & = & \overline{\ell} \gamma^{\mu}
\mathbf{T}_3 \ell - s^2_{\rm eff} \mathcal{J}^{\mu}_{\rm em} (\ell). 
\eea
Further, we denote the vertex functions for the \mbox{$\gamma WW$} and
\mbox{$ZWW$}~vertices obtained from the Lagrangian terms $\mathscr{L}_{\gamma
WW}$ and $\mathscr{L}_{ZWW}$, see~(\ref{eq-tgcsgamma}) and~(\ref{eq-tgcsz}),
by $\Gamma_{\gamma WW}$ and $\Gamma_{ZWW}$, respectively.  They include SM as
well as anomalous contributions, and no linear approximation in the~$h_i$ is
performed yet.  We have then for the sum of the amplitudes for photon and
$Z$~exchange in the $P_W$~scheme:
\bea
\label{eq-amplsum}
\mathcal{A}_{\gamma} + \mathcal{A}_Z & \propto & \mathcal{J}^{\mu}_{\rm em}
(\ell) \frac{1}{s} \Gamma_{\gamma WW} + G_{\rm NC} \mathcal{J}^{\mu}_{\rm NC}
(\ell) \frac{1}{s - m_Z^2} \Gamma_{ZWW}\nonumber\\
 & = & \mathcal{J}^{\mu}_{\rm em} (\ell) \frac{1}{s} \left. \Gamma_{\gamma WW}
 \right|_{\rm eff}\\
& & \mbox{} + G_{\rm NC}^{\rm SM} \left(\overline{\ell} \gamma^{\mu}
 \mathbf{T}_3 \ell - s_1^2 \mathcal{J}^{\mu}_{\rm em} (\ell)
 \right)\nonumber\\
& & \mbox{} \;\;\; \times \frac{1}{s - \left(m_Z^{\rm SM}\right)^2}
 \left. \Gamma_{ZWW}\right|_{\rm eff}, \nonumber
\eea
where we have defined
\be
G_{\rm NC}^{\rm SM} = \frac{1}{s_1 c_1},\msp \msp m_Z^{\rm SM} =
\frac{m_W}{c_1},
\ee
and the effective vertex functions
\bea
\label{eq-defeffphoton}
\left. \Gamma_{\gamma WW} \right|_{\rm eff} & = & \Gamma_{\gamma WW}\\
& & \mbox{} + \frac{s}{s - m_Z^2} G_{\rm NC} \left(s_1^2 - s_{\rm eff}^2\right)
\Gamma_{ZWW},\nonumber\\
\label{eq-defeffzboson}
\left. \Gamma_{ZWW}\right|_{\rm eff} & = & \frac{G_{\rm NC}}{G_{\rm NC}^{\rm
    SM}} \frac{s - \left(m_Z^{\rm SM}\right)^2}{s - m_Z^2} \Gamma_{ZWW}.
\eea
The squared c.m.~energy of the electron-positron system is denoted by~$s$.
From~(\ref{eq-amplsum}) we see that the sum of $\mathcal{A}_{\gamma}$ and
$\mathcal{A}_Z$ can be calculated from the diagrams in Fig.~\ref{fig:photon}
and diagrams~(a) and~(d) in Fig.~\ref{fig:zboson} if we use the vertex
functions \mbox{$\left. \Gamma_{\gamma WW} \right|_{\rm eff}$}
and~\mbox{$\left. \Gamma_{ZWW} \right|_{\rm eff}$} instead of~$\Gamma_{\gamma
WW}$ and~$\Gamma_{ZWW}$.  Expanding the coefficients of~$\Gamma_{ZWW}$
in~(\ref{eq-defeffphoton}) and~(\ref{eq-defeffzboson}) to linear order in the
$h_i$ we have, using~(\ref{eq-seffpwscheme}),
\bea
\left. \Gamma_{\gamma WW} \right|_{\rm eff} & = & \Gamma_{\gamma WW} -
\frac{s}{s - m_W^2/c_1^2} h_{W \mskip -3mu B} \Gamma_{ZWW},\\
\left. \Gamma_{ZWW}\right|_{\rm eff} & = & \bigg\{1 + \frac{s_1}{c_1}\left(1 +
  4 P(s)\right) h_{W \mskip -3mu B} + P(s) h_{\varphi}^{(3)}\bigg\}\nonumber\\
& & \mbox{} \;\;\; \times \Gamma_{ZWW}
\eea
with
\be
P(s) = \frac{m_W^2/2}{c_1^2 s - m_W^2}.
\ee
We can now think of \mbox{$\left. \Gamma_{\gamma WW} \right|_{\rm eff}$} and
\mbox{$\left. \Gamma_{ZWW} \right|_{\rm eff}$} as vertex functions emerging
from the Lagrangian terms~(\ref{eq-tgcsgamma}), (\ref{eq-tgcsz}) and
containing couplings \mbox{$\left. \Delta g_1^{\gamma}\right|_{\rm eff}$},
\mbox{$\left. \Delta g_1^Z\right|_{\rm eff}$}, etc.\ instead of \mbox{$\Delta
g_1^{\gamma}$}, \mbox{$\Delta g_1^Z$}, etc.  Taking into account the
additional factor of~\mbox{$(c_1/s_1)$} in the SM~couplings of~$\Gamma_{ZWW}$
compared to the SM~couplings of~$\Gamma_{\gamma WW}$, see~(\ref{eq-lagrhagi})
to~(\ref{eq-smtgcs}), we obtain to linear order in the~$h_i$
from~(\ref{eq-coup1a}) and~(\ref{eq-coup2a})
\bea
\label{eq-eff1}
\left. \Delta g_1^{\gamma}\right|_{\rm eff} & = & - \frac{c_1^3}{s_1}
\frac{2s}{m_W^2} P(s) h_{W \mskip -3mu B},\\
\label{eq-eff2}
\left. \Delta \kappa_{\gamma} \right|_{\rm eff} & = & - \frac{2 c_1}{s_1} P(s)
h_{W \mskip -3mu B},\\
\label{eq-eff3}
\left. \Delta g_1^Z\right|_{\rm eff} & = & \frac{s_1}{c_1} \left(1 + 4
  P(s)\right) h_{W \mskip -3mu B} + P(s) h_{\varphi}^{(3)},\\
\label{eq-eff4}
\left. \Delta \kappa_Z\right|_{\rm eff} & = & P(s) \left(\frac{4 s_1}{c_1}
    h_{W \mskip -3mu B} + h_{\varphi}^{(3)}\right).
\eea
With all other couplings \mbox{$\left. \lambda_{\gamma}\right|_{\rm eff}$},
\mbox{$\left. \lambda_Z\right|_{\rm eff}$}, etc.\ of the vertex functions
\mbox{$\left. \Gamma_{\gamma WW} \right|_{\rm eff}$} and
\mbox{$\left. \Gamma_{ZWW} \right|_{\rm eff}$} we drop the subscript~``eff''
and write $\lambda_{\gamma}$, $\lambda_Z$, etc.\ as usual since they are
related to the~$h_i$ as before according to~(\ref{eq-coup3a})
to~(\ref{eq-coup6a}).  In the high-energy limit \mbox{$s \gg m_W^2$} we obtain
from~(\ref{eq-eff1}) to~(\ref{eq-eff4})
\bea
\label{eq-eff1a}
\left. \Delta g_1^{\gamma}\right|_{\rm eff} & \approx & - \frac{c_1}{s_1} h_{W
  \mskip -3mu B},\\
\label{eq-eff2a}
\left. \Delta \kappa_{\gamma} \right|_{\rm eff} & \approx & 0,\\
\label{eq-eff3a}
\left. \Delta g_1^Z\right|_{\rm eff} & \approx & \frac{s_1}{c_1} h_{W
  \mskip -3mu B},\\
\label{eq-eff4a}
\left. \Delta \kappa_Z\right|_{\rm eff} & \approx & 0.
\eea
The effective couplings do therefore not depend on~$h_{\varphi}^{(3)}$ in this
limit.  We recall that three of the gauge relations in the $P_W$~scheme are
\bea
\Delta g_1^{\gamma} & = & 0,\\[.1cm]
\Delta g_1^Z & = & 0,\\
\label{eq-noeff}
\Delta \kappa_Z & = & \Delta g_1^Z - \frac{s_1^2}{c_1^2} \Delta
\kappa_{\gamma},\eea
see~(\ref{eq-gr1}), (\ref{eq-gr2}) with \mbox{$s_0 \rightarrow s_1$} and
  \mbox{$c_0 \rightarrow c_1$}, and~(\ref{eq-grg1z}).  Here, instead of these
  three relations we obtain two relations among the effective couplings
\bea
\label{eq-effrel1}
\left. \Delta g_1^{\gamma}\right|_{\rm eff} & = & c_1^2 \frac{s}{m_W^2}
\left. \Delta \kappa_{\gamma} \right|_{\rm eff},\\
\label{eq-effrel2}
\left. \Delta \kappa_Z\right|_{\rm eff} & = & \left. \Delta g_1^Z\right|_{\rm
  eff} - \frac{s_1^2}{c_1^2} \left. \Delta \kappa_{\gamma} \right|_{\rm eff}
  \left(-2 P(s)\right)^{-1}.
\eea
Notice the extra factor in the brackets in~(\ref{eq-effrel2}) compared to the
conventional relation~(\ref{eq-noeff}).  Instead of~(\ref{eq-effrel2}) one can
also choose a relation, whose coefficients are energy independent:
\be
\label{eq-effrel2a}
\left. \Delta \kappa_Z\right|_{\rm eff} = \left. \Delta g_1^Z\right|_{\rm
  eff} - \frac{s_1^2}{c_1^2} \left(\left. \Delta \kappa_{\gamma} \right|_{\rm
    eff} - \left. \Delta g_1^{\gamma}\right|_{\rm eff}\right).
\ee
However, not {\em both} gauge relations between the effective couplings
\mbox{$\left. \Delta g_1^{\gamma}\right|_{\rm eff}$}, \mbox{$\left. \Delta
\kappa_{\gamma} \right|_{\rm eff}$}, \mbox{$\left. \Delta g_1^Z\right|_{\rm
eff}$} and~\mbox{$\left. \Delta \kappa_Z\right|_{\rm eff}$} can be chosen with
energy independent coefficients.  This can be seen in the following way:
Assume that in addition to~(\ref{eq-effrel2a}) there is a gauge relation
\be
\label{eq-newgr}
A \left. \Delta g_1^{\gamma}\right|_{\rm eff} + B \left. \Delta
  g_1^Z\right|_{\rm eff} + C \left. \Delta \kappa_{\gamma} \right|_{\rm eff} +
D \left. \Delta \kappa_Z\right|_{\rm eff} = 0,
\ee
where $A$, $B$, $C$ and $D$ are constants.  In the limit \mbox{$s \gg m_W^2$},
cf.\ (\ref{eq-eff1a}) to~(\ref{eq-eff4a}), we obtain from~(\ref{eq-newgr})
\be
\label{eq-ABrel}
B s_1^2 = A c_1^2.
\ee
Now, assuming~(\ref{eq-newgr}) to be independent from~(\ref{eq-effrel2a}), we
can without loss of generality set~\mbox{$A = 0$}.  Due to~(\ref{eq-ABrel}) we
then have also~\mbox{$B = 0$}.  The relation~(\ref{eq-newgr}) is then a
relation solely between \mbox{$\left. \Delta \kappa_{\gamma} \right|_{\rm
eff}$} and~\mbox{$\left. \Delta \kappa_Z\right|_{\rm eff}$}, which is not
possible because these couplings are obviously independent,
see~(\ref{eq-eff2}) and~(\ref{eq-eff4}).  Thus no such
relation~(\ref{eq-newgr}) with energy independent coefficients exists.
Instead at least one gauge relation, e.g.~(\ref{eq-effrel1}), depends on~$s$.
To summarise we obtain the following gauge relations among the effective
couplings (as mentioned above for all but four couplings we drop the
subscript~``eff''):
\bea
\label{eq-gr1a}
\left. \Delta g_1^{\gamma}\right|_{\rm eff} & = & c_1^2 \frac{s}{m_W^2}
\left. \Delta \kappa_{\gamma} \right|_{\rm eff},\\
\label{eq-gr2a}
\left. \Delta \kappa_Z\right|_{\rm eff} & = & \left. \Delta g_1^Z\right|_{\rm
  eff} - \frac{s_1^2}{c_1^2} \left. \Delta \kappa_{\gamma} \right|_{\rm eff}
  \left(-2 P(s)\right)^{-1},\\
\label{eq-gr3a}
\lambda_Z & = & \lambda_{\gamma},\\
\label{eq-gr4a}
\tilde{\kappa}_{\gamma} & = & - \frac{c_1^2}{s_1^2} \tilde{\kappa}_Z,\\
\label{eq-gr5a}
\tilde{\lambda}_{\gamma} & = & \tilde{\lambda}_Z,\\
\label{eq-gr6a}
g_4^{\gamma} & = & g_4^Z \;\,=\;\, g_5^{\gamma} \;\,=\;\, g_5^Z \;\,=\;\, 0.
\eea
Instead of~(\ref{eq-gr2a}) one may take the relation~(\ref{eq-effrel2a}) with
energy independent coefficients.

Numerically we find from~(\ref{eq-coup3a}) to~(\ref{eq-coup6a}) that the
couplings $\lambda_Z$,\ldots, $g_5^Z$ are expressed as linear combinations of
the parameters~$h_i$ in the following way:
\begin{alignat}{3}
\label{eq-coup3b}
\lambda_Z & = 0.980 h_W,& \lambda_{\gamma} & = &&\;
0.980 h_W,\\[.1cm]
\label{eq-coup4b}
\tilde{\kappa}_Z & = - 0.544 h_{\tilde{W} \mskip -3mu B},&
\tilde{\kappa}_{\gamma} & = &&\; 1.84 h_{\tilde{W} \mskip -3mu B},\\[.1cm]
\label{eq-coup5b}
\tilde{\lambda}_Z & = 0.980 h_{\tilde{W}},&
\tilde{\lambda}_{\gamma} & = &&\; 0.980 h_{\tilde{W}},\\[.1cm]
\label{eq-coup6b}
g_4^{\gamma} & = g_4^Z = g_5^{\gamma} = g_5^Z = 0.&&&&
\end{alignat}
For \mbox{$\sqrt{s} = 500$}~GeV we further obtain with~(\ref{eq-eff1})
to~(\ref{eq-eff4})
\bea
\label{eq-eff1b}
\left. \Delta g_1^{\gamma}\right|_{\rm eff} & = & -1.90 h_{W \mskip -3mu B},\\
\label{eq-eff2b}
\left. \Delta \kappa_{\gamma} \right|_{\rm eff} & = & -0.064 h_{W \mskip -3mu
  B},\\
\label{eq-eff3b}
\left. \Delta g_1^Z\right|_{\rm eff} & = & 0.582 h_{W \mskip -3mu B} + 0.017
h_{\varphi}^{(3)},\\
\label{eq-eff4b}
\left. \Delta \kappa_Z\right|_{\rm eff} & = & 0.038 h_{W \mskip -3mu B} + 0.017
h_{\varphi}^{(3)}.
\eea
For \mbox{$\sqrt{s} = 800$}~GeV, we have instead of~(\ref{eq-eff1b})
to~(\ref{eq-eff4b})
\bea
\label{eq-eff1c}
\left. \Delta g_1^{\gamma}\right|_{\rm eff} & = & -1.86 h_{W \mskip -3mu B},\\
\label{eq-eff2c}
\left. \Delta \kappa_{\gamma} \right|_{\rm eff} & = & -0.024 h_{W \mskip -3mu
  B},\\
\label{eq-eff3c}
\left. \Delta g_1^Z\right|_{\rm eff} & = & 0.558 h_{W \mskip -3mu B} + 0.007
h_{\varphi}^{(3)},\\
\label{eq-eff4c}
\left. \Delta \kappa_Z\right|_{\rm eff} & = & 0.014 h_{W \mskip -3mu B} +
0.007 h_{\varphi}^{(3)}.
\eea
In the high-energy limit~\mbox{$s \gg m_W^2$} we obtain from~(\ref{eq-eff1a})
to~(\ref{eq-eff4a})
\bea
\label{eq-eff1d}
\left. \Delta g_1^{\gamma}\right|_{\rm eff} & \approx & -1.84 h_{W \mskip -3mu
  B},\\
\label{eq-eff2d}
\left. \Delta \kappa_{\gamma} \right|_{\rm eff} & \approx & 0,\\
\label{eq-eff3d}
\left. \Delta g_1^Z\right|_{\rm eff} & \approx & 0.544 h_{W \mskip -3mu B},\\
\label{eq-eff4d}
\left. \Delta \kappa_Z\right|_{\rm eff} & \approx & 0.
\eea
From the measurements of \mbox{$\left. \Delta g_1^{\gamma}\right|_{\rm eff}$},
\mbox{$\left. \Delta \kappa_{\gamma}\right|_{\rm eff}$},\ldots, $g_5^Z$ in the
reaction~\mbox{$e^+ e^- \rightarrow WW$} at a future LC,
see~\cite{Diehl:2002nj,Diehl:2003qz}, we can thus get bounds on~$h_W$, $h_{W
\mskip -3mu B}$, $h_{\varphi}^{(3)}$, $h_{\tilde{W}}$ and~$h_{\tilde{W} \mskip
-3mu B}$ if $s$ is not too large.  In the high-energy limit \mbox{$s \gg
m_W^2$} the $CP$~conserving coupling~$h_{\varphi}^{(3)}$ cannot be measured in
this way.

%%%%%%%%%%%%%%%%%%%%%%%%%%%%%%%%%%%%%%%%%%%%%%%%%%%%%%%%%%%%%%%%%

\subsection[Bounds from $e^+ e^- \rightarrow WW$ at a linear collider]{Bounds
  from \boldm{e^+ e^- \rightarrow WW} at a linear collider}
\label{ssec-boundsfromnlc}

In this section we discuss the reaction \mbox{$e^+ e^- \rightarrow WW$}, to be
measured at a future linear collider, in view of its sensitivity to the
anomalous couplings~$h_i$.  We assume unpolarised $e^+$ and $e^-$ beams and
standard expected values for the integrated
luminosities~\cite{Aguilar-Saavedra:2001rg,Ellis:1998wx} 500~fb$^{-1}$ at
\mbox{$\sqrt{s} = 500$}~GeV, 1~ab$^{-1}$ at \mbox{$\sqrt{s} = 800$}~GeV and
3~ab$^{-1}$ at~\mbox{$\sqrt{s} = 3$}~TeV.  We use the errors for all TGCs in
the parameterisation~(\ref{eq-lagrhagi}), as given for \mbox{$\sqrt{s} =
500$}~GeV and  \mbox{$\sqrt{s} = 800$}~GeV in Tabs.~5 and~9
of~\cite{Diehl:2003qz}, respectively, and take into account their correlations
(which are not listed there).  We further use the corresponding results
calculated for~\mbox{$\sqrt{s} = 3$}~TeV.  From these values we can extract
the errors obtainable for the~$h_i$ using~(\ref{eq-coup3b})
to~(\ref{eq-eff4c}) by conventional error propagation.  We give the errors and
correlations at c.m.~energies of 500~GeV, 800~GeV and 3~TeV for the
$CP$~conserving couplings in Tabs.~\ref{tab:500even} to~\ref{tab:3000even} and
for the $CP$~violating ones in Tab.~\ref{tab:nlcodd}.  The errors of~$h_W$,
$h_{W \mskip -3mu B}$, $h_{\tilde{W}}$ and~$h_{\tilde{W} \mskip -3mu B}$  at
500~GeV are considerably smaller than the one on~$h_{\varphi}^{(3)}$.  Notice
that $h_{\varphi}^{(3)}$ becomes unmeasurable in the high-energy limit,
see~(\ref{eq-eff1d}) to~(\ref{eq-eff4d}).  At \mbox{$\sqrt{s} = 3$}~TeV we
thus obtain no bound on~$h_{\varphi}^{(3)}$.  For all other measurable
couplings the errors become much smaller with rising energy.  Notice that the
error correlations decrease with rising energy and the four measurable
couplings are almost uncorrelated at~\mbox{$\sqrt{s} = 3$}~TeV.
\begin{table}
\centering 
\begin{tabular}{lrrrr}
 $h$ & $\delta h \times 10^3$ & $h_W$ & $h_{W \mskip -3mu B}$ &
 $h_{\varphi}^{(3)}$\\[.02cm]
\hline
 &&&&\\[-.32cm]
$h_W$ & 0.28 & 1 & 0.09 & $-0.26$\\
$h_{W \mskip -3mu B}$ & 0.32 && 1 & $-0.73$\\
$h_{\varphi}^{(3)}$ & 36.4 &&& 1 
\end{tabular}
\caption{\label{tab:500even}Errors in units of $10^{-3}$ and correlations of
  the $CP$~conserving couplings at c.m.~energy \mbox{$\sqrt{s} = 500$}~GeV.}
\end{table}
\begin{table}
\centering 
\begin{tabular}{lrrrr}
 $h$ & $\delta h \times 10^3$ & $h_W$ & $h_{W \mskip -3mu B}$ &
 $h_{\varphi}^{(3)}$\\[.02cm]
\hline
 &&&&\\[-.32cm]
$h_W$ & 0.12 & 1 & 0.08 & $-0.15$\\
$h_{W \mskip -3mu B}$ & 0.16 && 1 & $-0.79$\\
$h_{\varphi}^{(3)}$ & 53.7 &&& 1 
\end{tabular}
\caption{\label{tab:800even}Same as Tab.~\protect{\ref{tab:500even}} but for
  \mbox{$\sqrt{s} = 800$}~GeV.} 
\end{table}
\begin{table}
\centering 
\begin{tabular}{lrrr}
 $h$ & $\delta h \times 10^3$ & $h_W$ & $h_{W \mskip -3mu B}$\\[.02cm]
\hline
 &&&\\[-.32cm]
$h_W$ & 0.018 & 1 & $-0.004$ \\
$h_{W \mskip -3mu B}$ & 0.015 && 1
\end{tabular}
\caption{\label{tab:3000even}Errors in units of $10^{-3}$ and correlations of
  the $CP$~conserving couplings in the high-energy limit at c.m.~energy
  \mbox{$\sqrt{s} = 3$}~TeV.}
\end{table}
\begin{table}
\centering
\begin{tabular}{rrrr}
$\sqrt{s}$ & $\delta h_{\tilde{W}} \times 10^3$ & $\delta h_{\tilde{W}
  \mskip -3mu B} \times 10^3$ & corr.\\[0.07cm]
\hline
 &&&\\[-.32cm]
500~GeV & 0.28 & 2.2 & 17\%\\
800~GeV & 0.12 & 1.4 & 9\%\\
3~TeV & 0.018 & 0.77 & 2\%
\end{tabular}
\caption{\label{tab:nlcodd}Errors in units of $10^{-3}$ and correlations of
  the $CP$~violating couplings at different c.m.~energies.}
\end{table}

%%%%%%%%%%%%%%%%%%%%%%%%%%%%%%%%%%%%%%%%%%%%%%%%%%%%%%%%%%%%%%%%%
%%%%%%%%%%%%%%%%%%%%%%%%%%%%%%%%%%%%%%%%%%%%%%%%%%%%%%%%%%%%%%%%%

\section{Conclusions}
\label{sec-conc}
\setcounter{equation}{0}

We have analysed the phenomenology of the gauge-boson sector of an electroweak
locally \mbox{$SU(2)\times U(1)$} invariant effective Lagrangian.  In addition
to the SM Lagrangian we took into account anomalous coupling terms from the
ten operators of dimension six built either only from the SM gauge fields or
from the SM gauge fields combined with the SM-Higgs-doublet field.  We found
that after SSB some anomalous terms contribute to the diagonal and
off-diagonal kinetic terms of the neutral gauge bosons, to the mass terms of
the $W$ and the $Z$~bosons, and to the kinetic term of the Higgs boson.  This
made necessary to first identify the physical neutral gauge-boson fields as
linear combinations of the fields that originally occur in the Lagrangian, and
to renormalise the Higgs-boson field and the charged gauge-boson fields.  In
this way, in addition to the gauge-boson self-interactions, also the neutral-
and charged-current interactions were modified.  A careful discussion of
electroweak parameterisation schemes was given, see Tab.~\ref{tab:parameters}.
We have studied the impact of anomalous couplings onto LEP and SLC
observables.  For a large class of observables the anomalous effects only show
up through a modified effective leptonic weak mixing angle, see
Sect.~\ref{sec-lep}.  The functional dependence of these observables on the
effective mixing angle is the same as in the~SM.  Thus the discrepancy between
the predictions for this angle from hadronic and leptonic observables cannot
be obtained by non-zero anomalous couplings from our boson operators.  The
observables $\Gamma_Z$, $m_W$ and $\Gamma_W$, depend on the anomalous
couplings in a different way and therefore lead to further constraints.  From
all these observables we obtain bounds of order~$10^{-3}$ for the
dimensionless couplings $h_{W \mskip -3mu B}$ and~$h_{\varphi}^{(3)}$.  These
bounds depend on~$m_H$.

Turning then to the TGCs we found that in addition to the two couplings $h_{W
 \mskip -3mu B}$ and~$h_{\varphi}^{(3)}$ one more $CP$~conserving coupling,
 $h_W$, and the two $CP$~violating couplings $h_{\tilde{W}}$ and $h_{\tilde{W}
 \mskip -3mu B}$ modify the \mbox{$\gamma WW$} and \mbox{$ZWW$} vertices in
 the scheme~$P_Z$.  In the scheme~$P_W$ the triple-gauge-boson vertices are
 parameterised by one coupling less than in~$P_Z$, see
 Tabs.~\ref{tab:parameters} and~\ref{tab:vertices}.  In other words there is
 an additional gauge relation in the scheme~$P_W$.  However, both with~$P_Z$
 and with~$P_W$ some $CP$~conserving couplings also change the boson-fermion
 interactions.  For the specific reaction~\mbox{$e^+e^- \rightarrow WW$} and
 using $P_W$ we have defined effective TGCs such that all anomalous effects
 are absorbed into the effective three-gauge-boson vertices
 \mbox{$\left. \Gamma_{\gamma WW}\right|_{\rm eff}$}
 and~\mbox{$\left. \Gamma_{ZWW}\right|_{\rm eff}$}.  The anomalous
 gauge-boson-fermion interactions are thus fully taken into account here (in
 the approximation linear in the~$h_i$) though in the explicit calculation of
 the differential cross section everything apart from the TGCs is assumed to
 be SM like.  With the effective couplings one more parameter re-enters the
 differential cross-section in the scheme~$P_W$.  The gauge relations between
 the effective couplings are different from those between standard~TGCs.  At
 least one gauge relation contains the squared c.m.~energy~$s$ of the
 electron-positron system.

For the bounds derived from LEP2 data that includes various processes and not
only $W$-boson-pair production we have used $P_Z$ and only considered the
conventional TGCs.  This gives exact results for the $CP$~violating couplings,
but only approximate results for the $CP$~conserving ones, since we have
neglected the modified $W$~mass and boson-fermion interactions there.  For the
couplings $h_{W \mskip -3mu B}$ and $h_{\varphi}^{(3)}$ the direct LEP2
measurements do not give tighter bounds than the other LEP and SLC
observables.  However, we obtain in addition bounds on $h_W$, $h_{\tilde{W}}$
and $h_{\tilde{W} \mskip -3mu B}$ of order~0.1.

Our summary of the presently available information on the anomalous
couplings~$h_i$ is presented in Tabs.~\ref{tab:coup3} and~\ref{tab:coup4} and
in Fig.~\ref{fig:precdata}.  We find that the data is consistent with a light
Higgs boson, \mbox{$m_H = 120$}~GeV and practically vanishing anomalous
couplings.  But also a heavy Higgs boson, \mbox{$m_H \approx 500$}~GeV, is in
accordance with the present data if only small anomalous couplings~$h_{W
\mskip -3mu B}$ and~$h_{\varphi}^{(3)}$ of order~$10^{-3}$ are introduced in
the gauge-boson sector, see Fig.~\ref{fig:precdata}.  Moreover the data
prefers a value for~$h_W$ of~$-0.06$ over \mbox{$h_W = 0$} at the
\mbox{$2\sigma$}~level, see Tab.~\ref{tab:coup3}.  This may change if
radiative corrections are included in the relevant LEP2 analyses of~TGCs.

We have investigated in detail the effects of our effective Lagrangian on the
reaction \mbox{$e^+ e^- \rightarrow WW$} at a future~LC.  To this end we have
used the results obtained for solely TGCs in the most general parameterisation
for unpolarised beams and longitudinal polarisation~\cite{Diehl:2002nj} as
well as for transverse polarisation~\cite{Diehl:2003qz}.  These analyses have
been done with optimal observables and the derived constraints on the $h_i$
therefore give the optimal bounds that one can obtain in this reaction from
the normalised event distribution.  Here we have used the scheme~$P_W$ and our
technique with the effective vertices \mbox{$\left. \Gamma_{\gamma
WW}\right|_{\rm eff}$} and~\mbox{$\left. \Gamma_{ZWW}\right|_{\rm eff}$}.  For
most couplings the bounds obtainable with standard expected integrated
luminosities are \mbox{$\delta h_i$} around a few $10^{-4}$ to $10^{-3}$ at a
c.m.~energy \mbox{$\sqrt{s} = 500$}~GeV and are greatly improved with rising
energy.  Only one coupling, $h_{\varphi}^{(3)}$, is not measurable in the
high-energy limit.

Now we compare our results to the ones
of~\cite{Hagiwara:1992eh,Hagiwara:1993ck}.  The authors
of~\cite{Hagiwara:1992eh} have calculated at tree and one-loop level the
\mbox{$\gamma \gamma$}-, \mbox{$\gamma Z$}-, \mbox{$ZZ$}- and
\mbox{$WW$}-two-point functions as well as the vector-boson-fermion vertex
functions in an effective Lagrangian approach with two additional operators of
dimension six.  Thus they are more general in considering also loop effects
but in the present work we are more general in including more operators.

In the extensive work~\cite{Hagiwara:1993ck} a gauge-invariant effective
Lagrangian with dimension-six operators is considered.  There only $C$ and $P$
conserving operators are included.  The total set of operators that can be
constructed using the gauge fields and the Higgs field of the~SM is reduced by
discarding terms which are only total derivatives.  However, in contrast
to~\cite{Buchmuller:1985jz} and to our analysis here, the equations of motion
are not applied in the reduction of the number of operators since the authors
of~\cite{Hagiwara:1993ck} considered tree-level and one-loop effects.
Compared to this work we have studied here the tree-level effects of $C$, $P$
and \mbox{$CP$} conserving and violating operators.  We have also shown the
advantages and disadvantages of the two parameterisation schemes, $P_Z$ and
$P_W$, for the study of~TGCs.  Finally we have defined effective TGCs in the
scheme~$P_W$ which allow a direct comparison of the ELb, ELa and FF approaches
for \mbox{$e^+e^- \rightarrow WW$}.

An extensive study~\cite{Kuss:1997mf} has discussed the measurement of the
\mbox{$\gamma WW$} and \mbox{$ZWW$} couplings at the~LHC.  Using events with a
\mbox{$W^{\pm} Z$} (\mbox{$W^{\pm} \gamma$}) pair in the final state one is
sensitive only to the \mbox{$ZWW$} (\mbox{$\gamma WW$}) couplings in
Drell-Yan-type production, and therefore the two groups of couplings can be
measured separately.  Since our results are in the~ELb framework they cannot
be directly compared to those of~\cite{Kuss:1997mf} where merely anomalous
TGCs are assumed.  In~\cite{Kuss:1997mf} bounds on three~TGCs from events with
a \mbox{$W^{\pm} Z$} {\em or} a \mbox{$W^{\pm} \gamma$} pair are also computed
in a framework with a gauge-invariant effective Lagrangian.  However there,
too, effects from other vertices or propagators are not considered and
therefore also these results cannot be directly compared with ours.  For all
these reasons we conclude that a concise comparison of the sensitivity at
the~LHC with the bounds from a future~LC calculated in this paper requires a
full calculation of the processes there, which is beyond the scope of the
present work.  We should also note that the~TGCs studied for the~LHC
in~\cite{Kuss:1997mf} the \mbox{$ZWW$} and \mbox{$\gamma WW$} vertices are
studied for one~$W$ far off-shell, the other~$W$ and the~$Z$ and~$\gamma$
on-shell.  In our~LC study the two~$W$s are on-shell, the~$Z$ and~$\gamma$ far
off-shell.  We see that there is nice complementarity of the LHC and LC
possibilities.

Coming back to the results of our present paper we note that the Giga-$Z$ mode
at TESLA, see Sect.~5.1.4 of~\cite{Monig:2003cm}, will be particularly
interesting to accurately measure~$h_{W \mskip -3mu B}$ and
~$h_{\varphi}^{(3)}$.  A measurement at the $Z$~pole with an event rate that
is about 100~times that of LEP1, should in essence reduce the
errors~\mbox{$\delta h$} given in Tab.~\ref{tab:coup1} by a factor~10.  Thus
$h_{W \mskip -3mu B}$ and~$h_{\varphi}^{(3)}$ can then be measured with an
accuracy of some~$10^{-4}$.  However, systematical errors can become more
important there~\cite{Monig:2003ze}.

A very interesting opportunity for the exploration of the electroweak
gauge-boson sector is the measurement of the differential cross section of
\mbox{$\gamma \gamma \rightarrow WW$} at a photon collider, which we shall
explore in a future work~\cite{photon-collider}.  Here two new coupling
combinations can be determined that cannot be measured with the other options
that we have considered.

We have seen that experiments performed in the past as well as the
\mbox{Giga-$Z$}, the \mbox{$e^+e^-$} and the \mbox{$\gamma \gamma$} options at
a future LC all provide and will provide useful and complementary information
on the gauge-boson sector.  At present a non-zero value is preferred for~$h_W$
at the \mbox{$2\sigma$}~level, while small~$h_{W \mskip -3mu B}$
and~$h_{\varphi}^{(3)}$ can make a heavy Standard Model Higgs boson with
\mbox{$m_H \approx 500$}~GeV compatible with the data.  The bounds on the
$CP$~conserving anomalous couplings depend on the mass of the Higgs boson.
Until the Higgs boson is found the bounds on these couplings can therefore
only be given as a function of~$m_H$.  If a Higgs boson is discovered at the
LHC the constraints on the $CP$~conserving couplings from LEP and SLC
observables can be precisely stated.  The present bounds on the~$CP$~violating
couplings are rather loose.  In the future, with data from all three mentioned
linear collider modes seven out of ten anomalous coupling combinations can be
measured.  Our study in this paper and the one to follow on the
reaction~\mbox{$\gamma \gamma \rightarrow WW$} should make it clear that
exploring the electroweak gauge structure needs a comprehensive study at a
future linear collider where all running modes are needed and will reveal
interesting complementary aspects.

\section*{Acknowledgements}

The authors are very grateful to W.~Buchm\"uller, A.~Denner, M.~Diehl,
W.~Kilian, W.~Menges, K.~M\"onig and T.~Ohl for useful discussions.  Special
thanks to M.~Diehl for reading a draft.  For remarks on the first version of
this paper we are grateful to B.~Grzadkowski and V.~Spanos, and especially to
A.~Denner and M.~Spira.  This work was supported by the German
Bundesministerium f\"ur Bildung und Forschung, BMBF project no.~05HT4VHA/0,
and the Deutsche Forschungsgemeinschaft through the Graduiertenkolleg
``Physikalische Systeme mit vielen Freiheitsgraden''.

%%%%%%%%%%%%%%%%%%%%%%%%%%%%%%%%%%%%%%%%%%%%%%%%%%%%%%%%%%%%%%%%%
%%%%%%%%%%%%%%%%%%%%%%%%%%%%%%%%%%%%%%%%%%%%%%%%%%%%%%%%%%%%%%%%%

\end{document}